\documentclass[twocolumn]{aastex63}

\usepackage{graphicx}	
\usepackage{amsmath}	
\usepackage{amssymb}	
\usepackage{xspace}

\newcommand{\Ha}{H$\alpha$\xspace}

\newcommand{\hdue}{$\rm H_{2}$\xspace}
\newcommand{\Msun}{$\rm M_\odot$\xspace}
\newcommand{\sighdue}{$\Sigma_{H_2}$\xspace}
\newcommand{\kms}{$\rm km \, s^{-1}$\xspace}

\received{}
\revised{}
\accepted{\today}
\submitjournal{ApJ}

\shorttitle{ALMA view on JW100}
\shortauthors{Moretti et al.}

\graphicspath{{./}{figures/}}

\begin{document}

\title{GASP. XXII The molecular gas content of the JW100 jellyfish galaxy at z$\sim$0.05: does ram pressure promote molecular gas formation?}

\correspondingauthor{Alessia Moretti}
\email{alessia.moretti@inaf.it}

\author[0000-0002-1688-482X]{Alessia Moretti}
\affiliation{INAF-Padova Astronomical Observatory, Vicolo dell'Osservatorio 5, I-35122 Padova, Italy}

\author[0000-0001-9143-6026]{Rosita Paladino}
\affiliation{INAF-Istituto di Radioastronomia, via P. Gobetti 101, I-40129 Bologna, Italy}

\author[0000-0001-8751-8360]{Bianca M. Poggianti}
\affiliation{INAF-Padova Astronomical Observatory, Vicolo dell'Osservatorio 5, I-35122 Padova, Italy}

\author[0000-0001-5965-252X]{Paolo Serra}
\affiliation{INAF-Cagliari Astronomical Observatory, Via della Scienza 5, I-09047 Selargius (CA), Italy}

\author[0000-0003-2076-6065]{Elke Roediger}
\affiliation{E.A. Milne Centre for Astrophysics, Department of Physics and Mathematics, University of Hull, Hull, HU6 7RX, UK}

\author[0000-0002-7296-9780]{Marco Gullieuszik}
\affiliation{INAF-Padova Astronomical Observatory, Vicolo dell'Osservatorio 5, I-35122 Padova, Italy}

\author[0000-0002-8238-9210]{Neven Tomi\v{c}i\'{c}}
\affiliation{INAF-Padova Astronomical Observatory, Vicolo dell'Osservatorio 5, I-35122 Padova, Italy}

\author[0000-0002-3585-866X]{Mario Radovich}
\affiliation{INAF-Padova Astronomical Observatory, Vicolo dell'Osservatorio 5, I-35122 Padova, Italy}

\author[0000-0003-0980-1499]{Benedetta Vulcani}
\affiliation{INAF-Padova Astronomical Observatory, Vicolo dell'Osservatorio 5, I-35122 Padova, Italy}

\author[0000-0003-2150-1130]{Yara L. Jaff\'e}
\affiliation{Instituto de Fisica y Astronomia, Universidad de Valparaiso, Avda. Gran Bretana 1111 Valparaiso, Chile,}

\author[0000-0002-7042-1965]{Jacopo Fritz}
\affiliation{Instituto de Radioastronomia y Astrofisica, UNAM, Campus Morelia, AP 3-72, CP 58089, Mexico}

\author[0000-0002-4158-6496]{Daniela Bettoni}
\affiliation{INAF-Padova Astronomical Observatory, Vicolo dell'Osservatorio 5, I-35122 Padova, Italy}

\author[0000-0003-0231-3249]{Mpati Ramatsoku}
\affiliation{Department of Physics and Electronics, Rhodes University, PO Box 94, Makhanda, 6140, South Africa}
\affiliation{South African Radio Astronomy Observatory, 2 Fir Street, Black River Park, Observatory, Cape Town, 7405, South Africa}
\affiliation{INAF-Cagliari Astronomical Observatory, Via della Scienza 5, I-09047 Selargius (CA), Italy}

\author[0000-0001-5840-9835]{Anna Wolter}
\affiliation{INAF-Brera Astronomical Observatory, via Brera 28, I-20121 Milano, Italy}

\begin{abstract}
Within the GASP survey, aimed at studying the effect of the ram-pressure stripping on the star formation quenching in cluster galaxies, we analyze here ALMA observations of the jellyfish galaxy JW100.
We find an unexpected large amount of molecular gas  ($\sim 2.5 \times 10^{10}$ \Msun), 30\% of which is located in the stripped gas tail out to $\sim$35 kpc from the galaxy center.
The overall kinematics of molecular gas is similar to the one shown by the ionized gas, but for clear signatures of double components along the stripping direction detected only out to 2 kpc from the disk. 
The line ratio $r_{21}$ has a clumpy distribution and in the tail can reach large values ($\geq 1$), while its average value is low (0.58 with a 0.15 dispersion).
All these evidence strongly suggest that the molecular gas in the tail is newly born from stripped HI gas or newly condensed from stripped diffuse molecular gas.
The analysis of interferometric data at different scales reveals that a significant fraction ($\sim 40\%$) of the molecular gas is extended over large scales ($\geq 8$ kpc) in the disk, and this fraction becomes predominant in the tail ($\sim 70\%$).
By comparing the molecular gas surface density with the star formation rate surface density derived from the \Ha emission from MUSE data, we find that the depletion time on 1 kpc scale is particularly large ($5-10$ Gyr) both within the ram-pressure disturbed region in the stellar disk, and in the complexes along the tail.

\end{abstract}


\section{Introduction} \label{sec:intro}
Numerous studies in the recent years have started analyzing with unprecedented spatial resolution the ionized gas emission coming from cluster spiral galaxies showing extended gaseous tails, aiming at probing the effect of cluster-induced interactions on galaxy evolution.
The main mechanism acting on a galaxy gas without altering its stellar component is ram-pressure stripping \citep{GG72}, which has been demonstrated indeed to explain fairly well the observed data, despite its simplicity \citep{Jaffe2015,gaspIX}.
The availability of Integral Field Units (IFU), such as the MUSE spectrograph at VLT \citep{Bacon+2010}, has enormously boosted this field of research, allowing to map  the entire extent of galaxy tails, either with mosaic pointings for nearby galaxies \citep{Fumagalli2014,Fossati2016,Consolandi+2017} or with single pointings for more distant galaxies \citep{Merluzzi2013,gaspI,gaspII,gaspIII,gaspIV,gaspV}.
However, to completely characterize the ram-pressure stripping phenomenon, a complete mapping of all the other gas phases is mandatory:
X-ray tails have been detected so far only in few nearby ram-pressure stripped galaxies \citep{Machacek+2005,Sun2007,Sun2010}, whereas HI tails have been found in nearby cluster galaxies \citep{Kenney2004,Oosterloo2005,Chung2009,Abramson+2011,Scott+2018}, and also more recently in the GASP JO206 galaxy \citep{gaspXVII}, located at z$\sim 0.05$, where it has been possible to associate a long HI tail to the \Ha one. 

Cold molecular gas detection in the tail of nearby gas stripped spiral galaxies is more recent \citep{Jachym2014,Verdugo2015,Jachym2017,Lee+2017,Lee+2018,gaspX}, and was obtained making use of single dish telescopes, with a limited spatial resolution.
To date, the ESO137-001 galaxy in the Norma cluster is the only one that has been analyzed using ALMA band 6 data \citep{Jachym+2019}. These observations have allowed to map the cold gas through the molecular CO(2-1) transition on sub-kpc scales for the first time in a gas stripped tail.

In \cite{gaspX} we have started a campaign devoted to the study of the molecular gas content of gas stripped galaxies belonging to the GASP survey \footnote{https://web.oapd.inaf.it/gasp/} \citep{gaspI}, that aims at studying the effects of environmental interactions on nearby ($z\sim0.05$) cluster galaxies.
While the main survey is based on a VLT MUSE Large Program (GAs Stripping Phenomena in galaxies with MUSE, P. I. B. Poggianti) that traces the ionized gas components, complementary datasets at different wavelengths are being collected and have started offering a clear view on all the connected gas phases \citep[see also][]{gaspXVII,George+2018,deb19}.

In particular, in \cite{gaspX} we observed with the APEX telescope 4 GASP galaxies and detected molecular gas both in the galaxy disks and in correspondence of the stripped ionised gas tails. 
However, while the overall total gas masses are reliable, the APEX beam has a size of $\sim$28 \arcsec at the observed frequency of CO(2-1), and therefore 
does not allow to study the cold gas distribution on small scales.

In order to understand where the cold molecular gas is located and what is the origin of this gas, we have therefore collected ALMA interferometric data at $\sim$1 kpc resolution for the same galaxies, and we show in this paper the results we have obtained so far by studying the most massive galaxy in the sample, JW100.

Sec. \ref{sec:data}, is dedicated to the ALMA data description of the  observations of two carbon monoxide transitions, CO(2-1) and CO(1-0), while Sec. \ref{sec:analysis} shows the derivation of the molecular gas morphology and kinematics.
Given the availability of both Band 3 and Band 6 observations, we can also
quantify the line temperature ratio, $r_{21}$, in the entire field covered by both CO(2-1) and CO(1-0) observations, as we describe in detail in Sec. \ref{sec:r21}.
Sec. \ref{sec:mass} contains our estimates of the cold gas mass found in and around JW100, as well as the comparison between the interferometric and single dish masses. 
The comparison with MUSE data has finally allowed us for the first time to derive the Star Formation Efficiency (SFE) resolved on $\sim$1 kpc scale both in the disk and in the tail of this ram-pressure stripped galaxy, that is discussed in Sec. \ref{sec:sfe}.
A summary of our findings and the conclusions are in Sec. \ref{sec:conclusions}.
Throughout this paper we will make use of the standard cosmology $H_0 = 70 \, \rm km \, s^{-1} \, Mpc^{-1}$, ${\Omega}_M=0.3$
and ${\Omega}_{\Lambda}=0.7$, which yields 1$^{\prime\prime}$=1.071 kpc at the cluster redshift (z=0.055).
The galaxy redshift is, instead, z=0.06189.
As in the other GASP papers, our stellar masses are calculated adopting a \cite{Chabrier2003} IMF.

\section{Observations and data reduction}\label{sec:data}

In this paper we explore the molecular gas distribution of the ram pressure stripped galaxy JW100, located in the cluster A2626.
We have classified it as a jellyfish galaxy, as its \Ha emitting tail (extending out to at least $\sim$ 35 kpc) is longer than the stellar disk extension. Among the GASP targeted galaxies, most of those exhibiting such long tails also possess a central AGN \citep{poggianti2017}, as does JW100 (see also \citealt{gaspXIX}).

The MUSE data analysis has been performed as described in \citealt{gaspI}, and we just recall here that the galaxy has a mass of $\sim 3\times 10^{11}$ \Msun \citep{gaspX} and lies at a projected distance of $\sim 86$ kpc (0.05 in terms of R$_{200}$) from the cluster center.
The hosting cluster is located at z$\sim$0.055, and has a velocity dispersion of $\sim 650$ \kms \citep{Biviano+2017}. 
The Intra Cluster Medium (ICM) temperature is $\sim 3.5$ keV \citep{Ignesti+2018}.
The position of the galaxy in the cluster, together with its projected velocity compared with the cluster velocity dispersion ($v/\sigma_{cl}\sim 2.6$) confirm that this galaxy is located where ram-pressure should be most efficient, i.e. in the peak stripping region in the phase-space diagram, close to first pericentric passage \citep{gaspIX}.
A complete multiwavelength dataset has been collected for this galaxy (X, UV, optical, sub-mm), and has led to the careful analysis of its complex baryon cycling (Poggianti et al., accepted).

The ALMA observations of JW100 have been taken during Cycle 5 (project 2017.1.00496.S), using Band 3 and Band 6 to observe the CO(1-0) and (2-1) transitions, respectively. 
Both bands have been used in dual polarization mode, centering one spectral window (spw) on the redshifted frequency of the CO transition, 108.645 GHz for CO(1-0) and 217.294 GHz for CO(2-1), and setting other three spws to observe the continuum.
The spws centered on the CO lines include 1920 channels, and provide a spectral resolution of 0.976 MHz, corresponding to a velocity resolution of 3.1 and 1.5 \kms, respectively. 

The ALMA field of view at the average frequencies of the observations is 59.7 arcsec and 27.2 arcsec, in Band 3 and Band 6 respectively. To cover with homogeneous sensitivity an area of 57 $\times$ 62 arcsec, necessary to include the main galaxy and the tail, we set mosaics of 7 and 23 pointings, in Band 3 and Band 6, respectively.
The 12m array Band 3 observations were taken in 1 session, 40 min time on source, observing each pointing of the mosaic for 5.8 min. Band 6 observations have been taken in 2 sessions, for a total time on source of 96 min, observing each pointing for 4.2 min. 

A compact configuration of the array with a number of antennas ranging between 43 and 45 was used in both Bands. 
Adopting as a proxy for the longest baseline the 80th percentile of the uv distance and for the  shortest baseline the 5th percentile of the uv distance to calculate the maximum recoverable scale (eq. 7.4 and 7.7 in the ALMA Technical Handbook\footnote{https://almascience.eso.org/documents-and-tools/latest/alma-technical-handbook}) the actual configurations used provide a resolution and maximum recoverable scale of 0.98 and 13.38 arcsec, respectively, in Band 3, and of 0.7 and 7.5 arcsec, in Band 6.

Additional observations with the 7m ACA array have been obtained in Band 6 to increase the maximum recoverable scale. The primary beam of the ACA at the average frequency of the Band 6 observations is 46.6 arcsec, a 7 points mosaic has been used to cover the same area observed with the 12-m array.
Each pointing has been observed for $\sim$ 1 hr. A number of antennas ranging between 9 and 11 was used, with a 8.9 m shortest baseline, providing a maximum recoverable scale of $\sim$ 18 arcsec.

The 12m array and ACA datasets have been independently calibrated using the ALMA pipeline (version Pipeline-CASA51-P2-B). 
The calibrated data have been further analyzed with the CASA software (version 5.4.0-7; \citealt{CASA}).  
The continuum has been subtracted from the datasets independently using the CASA task {\it{uvcontsub}}. A linear fit of the continuum in the visually selected line-free channels of the spw centered on the line emission has been calculated in the visibility plane and subtracted from the full spw.

The flux density scale has been fixed using observations of J0006-0623 (for ACA observations in Band 6), and J2253+1608 for the 12m-array observations both in Band 3 and Band 6. We can estimate an uncertainty in the flux measured to be of the order of 10$\%$, including the calibration and clean errors. 
We checked the accuracy of the ALMA and ACA relative flux scales by comparing the flux on the overlapping spatial scales between the two arrays, and found that they are consistent within 10\%, which is the calibration uncertainty.
The 12m and ACA visibilities, calibrated and continuum subtracted, have been concatenated and used for the deconvolution.

The images of line and continuum emissions in Band 3 and Band 6 have been obtained using the CASA task {\it{tclean}}. Continuum images have been obtained using the three additional spws observed, covering a bandwidth of 5.25 GHz using Briggs weighting with a robust value of 0.5.
In Band 3 there is no continuum emission, while in Band 6 a peak, with S/N 15, located at RA 23:36:26 DEC 21:08:54.8 is detected.

The datacubes have been obtained smoothing the spectral resolution to 20 km/s. The channel velocities were computed in the LSRK velocity frame (radio convention) with a zero-point corresponding to the redshifted frequencies of the observed CO transitions (108.644GHz for the (1-0) and 217.258 for the (2-1)).
The continuum subtracted dirty cubes were cleaned in regions of line emission initially identified automatically (using the {\it automask} parameter in tclean), and further refined interactively to better account for possible faint residual emission.
The weighting used for the line images has been natural, to enhance the sensitivity. 
The achieved rms has been measured in line free channels of the cubes.

For both continuum and line images, the resulting  properties:  synthesized beams, peaks, and rms, are reported in Table \ref{tab:images}. 

\begin{table}
\begin{center}
\caption{Properties of Band 3 and Band 6 ALMA images: 
synthesized beam ($\theta_{maj}$, $\theta_{min}$, and PA), rms and peak of cleaned images.
Band 6 properties refer to the image obtained combining 12m and ACA data.}\label{tab:images}
\begin{tabular}{|l|c|c|c|c|c|}
\hline
& $\theta_{maj}$   &$\theta_{min}$   & PA  & rms  & Peak \\ 
&arcsec  &arcsec   & deg  & Jy/beam  & Jy/beam \\ 
\hline
&\multicolumn{5}{c|}{Band 3}\\
\hline
Cont  & 1.98  & 1.72  & -1.03  & 1.8$\cdot 10^{-5}$  &  $--$              \\ 
\hline
CO(1-0)       & 2     & 1.7   & 8.3    & 9$\cdot 10^{-4}$    & 2.5$\cdot 10^{-2}$  \\ 
\hline\hline
&\multicolumn{5}{c|}{Band 6}\\
\hline
Cont    & 1.35  & 1.08  & 33.7   &  4.4$\cdot 10^{-5}$ &  6.7 $\cdot 10^{-4}$ \\ 
\hline
CO(2-1)       & 1.4  &  1.12 & 33.04  & 8$\cdot 10^{-4}$     &  4.2$\cdot 10^{-2}$ \\ 
\hline

\end{tabular}
\end{center}
\end{table}
In the following analysis Band 6 data used are obtained by combining 12m and ACA data.

\section{Data analysis}\label{sec:analysis}

\subsection{Molecular gas morphology}\label{sec:moments}
Starting from the cleaned datacubes at 20 \kms resolution, we constructed a detection mask using the {\sc SoFiA} software \citep{sofia}.
In particular, we used SoFiA to: remove spatial and spectral noise variations from the datacubes; convolve the datacubes with several 3D smoothing kernels and, for each kernel, apply a detection threshold; merge detected voxels into 3D objects and reject false detections based on their integrated signal to noise and reliability calculated as \cite{Serra+2012}.

\begin{figure*}
    \centering
    \includegraphics[width=0.95\textwidth]{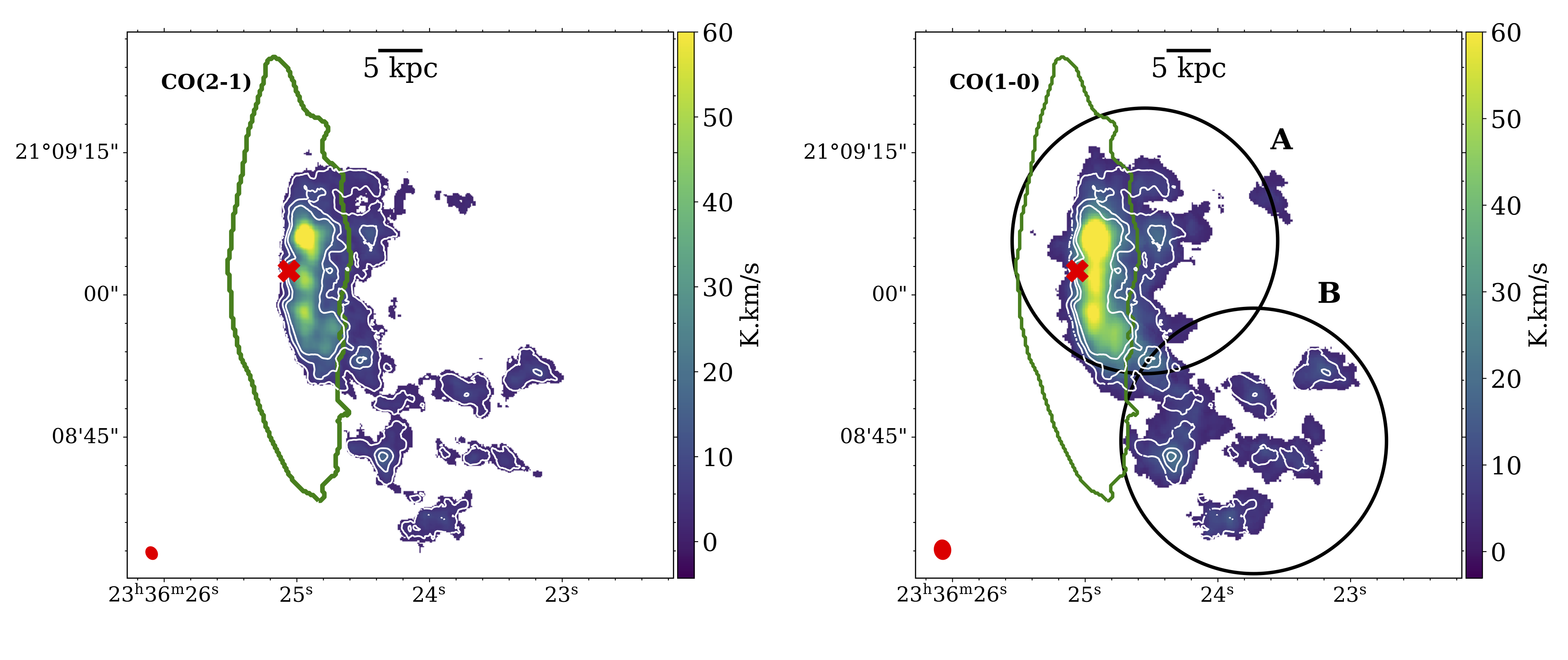}
    \includegraphics[width=0.95\textwidth]{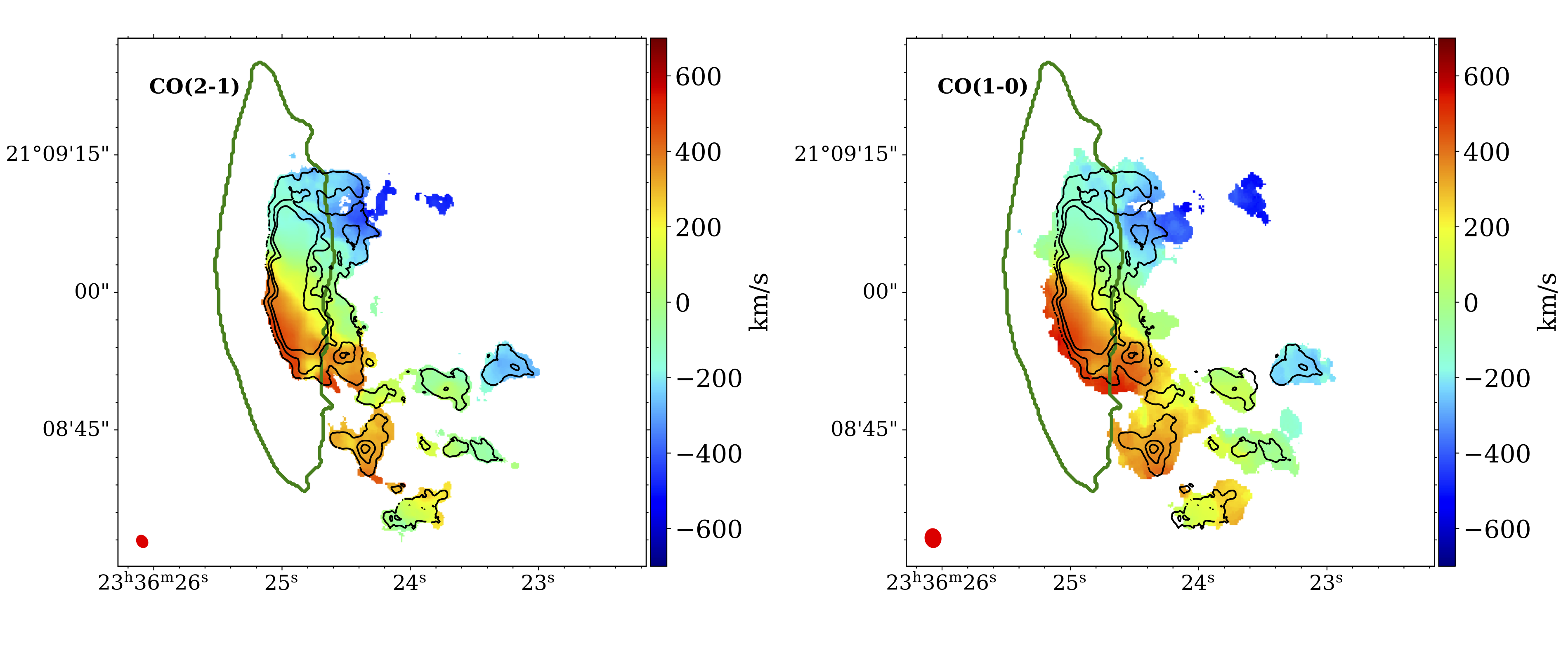}
    \includegraphics[width=0.95\textwidth]{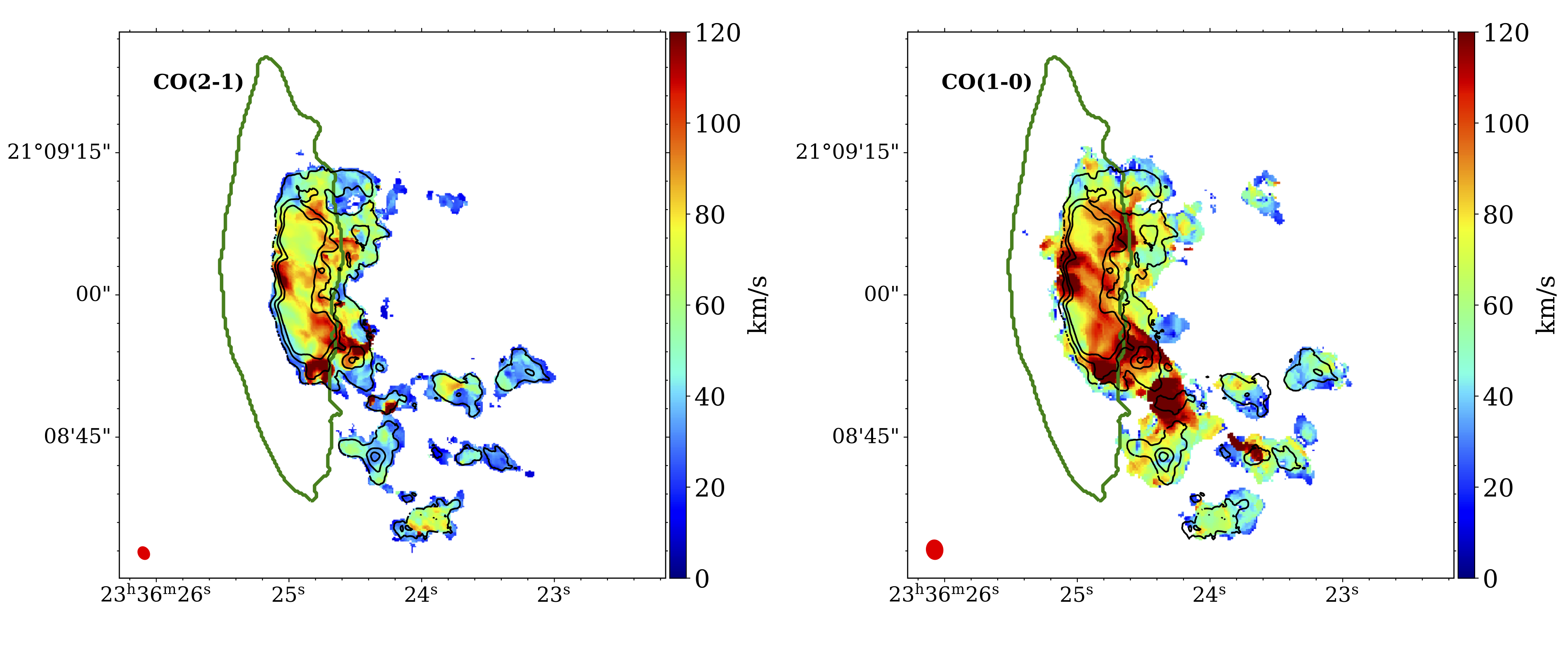}
    \caption{(Upper panels) CO(2-1) and CO(1-0) moments-zero from ALMA data, from 0 to 120 K.km/s.
    Beam sizes 
    are drawn in red in the bottom left corner. The red cross shows the position of the galaxy center. The two big black circles on the right panel are the original APEX pointings from \citealt{gaspX}. 
    (Middle, lower panels) CO(2-1) and CO(1-0) first moments and velocity dispersions (in \kms) from ALMA data. White contours (in the upper panels) and black contours in the others trace the CO(2-1) at 1, 3 and 5 moment zero RMS. In all panels the green contour shows the extent of the stellar disk from MUSE data. {In all panels north is up and east is left.}}
    \label{fig:Moments}
\end{figure*}

The first three moment maps are shown in Fig.~\ref{fig:Moments}. 
The upper panels  shows the moment-zero, i. e. the integrated line flux along the entire frequency range within the SoFiA masks, both for the CO(2-1) and for the CO(1-0).

The superimposed green contour shows the stellar disk extent derived from the most external ($\sim 1.5 \sigma$ above background) isophote of the continuum MUSE light under \Ha (as in Poggianti et al., accepted)
and in black, only in the CO(1-0) map, the two regions A and B covered by our APEX observations \citep{gaspX}. 
The temperature scale goes from 0 to 60 K.km/s. The beam size is drawn in red in the lower left corner of each map.
White contours superimposed in both panels are at 1, 3 and 5 moment zero RMS (defined as $RMS=RMS(cube) \times \sqrt{N_{chan}} \times \Delta_v)$).

Considering significant every detection of the line emission spread over 5 channels, i.e. with a linewidth of $\sim$ 100 \kms, 
the minimum measurable mass that we obtain using the relation given in Eqn. \ref{eqn:wk_co21}, that we will describe in Sec.~\ref{sec:mass}, is $\sim 1 \times 10^7$ \Msun.

The intensity weighted velocity (moment 1) and dispersion of velocity (moment 2) are shown in the middle and lower panels of Fig.~\ref{fig:Moments}, respectively, where the CO(2-1) contours are shown in black to guide the eye.

The CO emission appears elongated as the galaxy disk, but it is clearly less extended and displaced towards the west side of the disk, in the same direction of the tail of ionised gas revealed by MUSE and, therefore, of the ram-pressure wind.
The peak of the CO emission (both CO(2-1) and CO(1-0)) is located north-west with respect to the galaxy nucleus (shown as a red cross in the top panels of Fig.~\ref{fig:Moments}), at a distance of $\sim 5$ kpc. 
The peak in correspondence of the galaxy center seems to host less CO.
A similar missing flux in the central kpc has been observed also in the Andromeda galaxy \citep{Nieten+2006}, possibly originating from a satellite passage and in M51\citep{Schinnerer+2013}.
Ring-like structures have been found in a few galaxies belonging to the HERACLES survey \citep{Leroy2009} and in the NGC613 galaxy \citep{Miyamoto+2017} where the circum-nuclear disk also shows a very low star formation efficiency. 

CO emission is then present in the south-west tail in discrete regions with detected sizes up to $\sim 5$ kpc, confirming the APEX detection by \citet{gaspX}.

The extent of each individual region is clearly much bigger than the typical size of Milky Way Giant Molecular Clouds \citep{HeyerDame2015}, but our resolution is such that we can not exclude that each of them could be made of different unresolved regions.
It is remarkable, in any case, that ALMA data confirm the presence of CO emission at such large distances (up to $\sim$ 35 kpc) from the JW100 main body.

The distribution of CO(2-1) and CO(1-0) is very similar, but not entirely coincident, due to the larger extent of the CO(1-0), known to trace a more diffuse emission. 

\begin{figure*}
    \centering
    \includegraphics[width=0.9\textwidth]{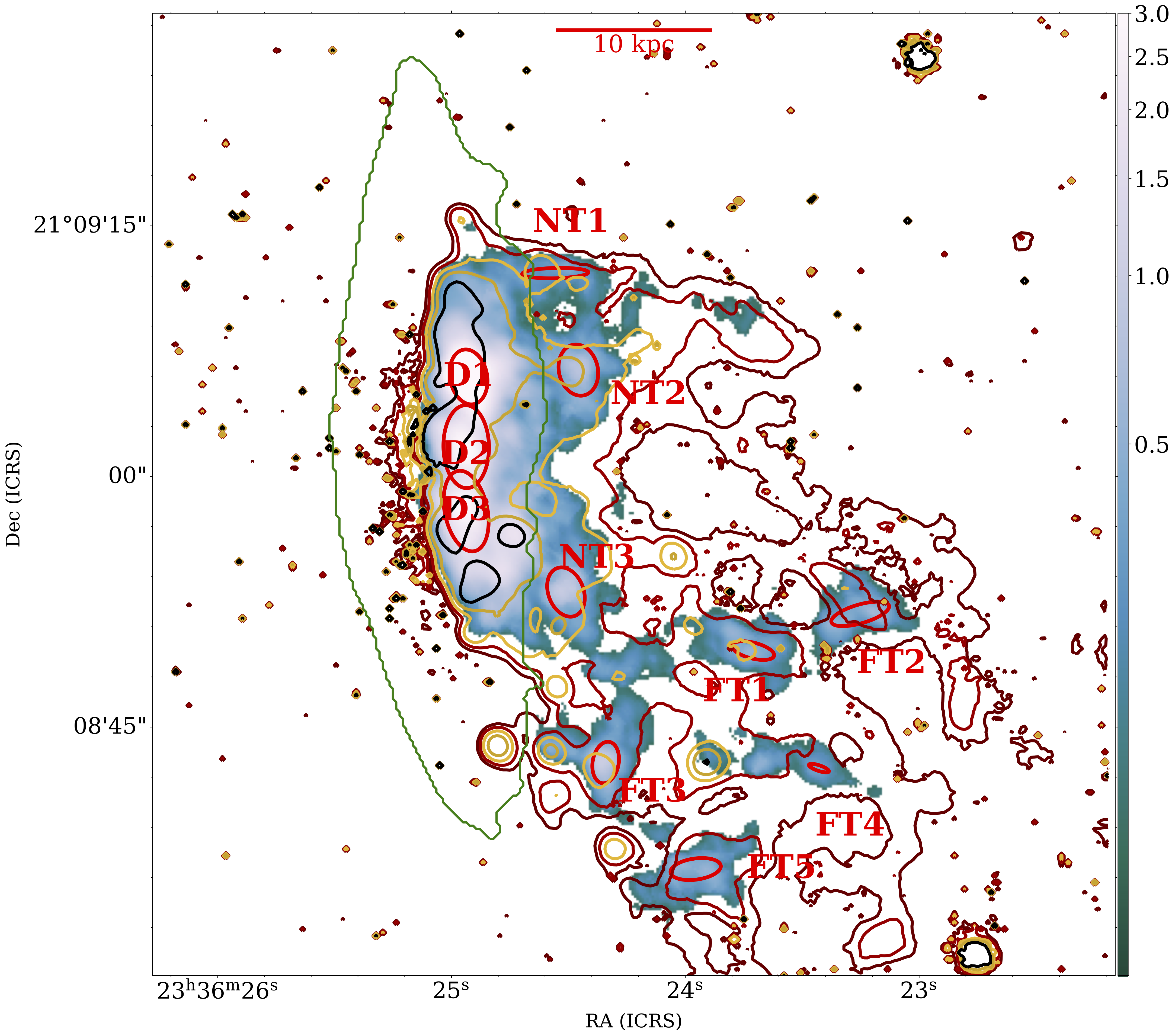}
    \caption{CO(2-1) emission (in Jy/beam.km/s) map of JW100 with superimposed the \Ha contours from MUSE (in different colors, at $2\times10^{-17}$,  $4\times10^{-17}$,  $8\times10^{-17}$,  $1.6\times10^{-16}$ and  $3.2\times10^{-15}$ erg/cm$^2$/s/arcsec$^2$) and the regions analyzed in this paper in red (label D1 to D3 for the disk regions, NT1 to NT3 for the clumps close to the disk and FT1 to FT5 for the clumps in the far tail). The green contour shows the extent of the stellar disk.}
    \label{fig:map}
\end{figure*}

In order to clearly identify the emitting regions, we fitted a 2d gaussian profile on each of them in the zero moment CO(2-1) map, shown again with a larger size in Fig.~\ref{fig:map}. 
This has allowed us to define a number of regions both within the galaxy disk, following the strongest CO concentration (regions D1, D2, D3 in Fig.~\ref{fig:map}), and the emissions close to the disk itself, but slightly displaced (regions NT1, NT2, and NT3 in Fig.~\ref{fig:map}), as well as the farthest clumps along the tail (regions FT1, FT2, FT3, FT4 and FT5 in  Fig.~\ref{fig:map}). 
The shape, orientation and size (gaussian FWHM) of each region are shown in Fig.~\ref{fig:map} (red ellipses).

The \Ha emission as derived from MUSE data, is shown with colored contours in Fig.~\ref{fig:map}. This emission remarkably traces the CO emission, as expected. 
However, there is one striking difference that is worth mentioning: 
the northern tail of \Ha has an overlapping CO emission with significant flux only within $\sim 15$ kpc from the galaxy center while its south-west extending tail is filled with knots of CO, with elongated shapes.
There are also \Ha emitting regions that do not have a molecular gas counterpart, especially in the south-west edge of the distribution but also close to the NT3 region. They are better discussed in Poggianti et al., accepted.
The two \Ha concentrations seen north-west and south-west are, instead, foreground stars.

ALMA data reveal therefore the presence of large clumps
of molecular gas well beyond the extent of the galaxy disk, thus posing an important challenge to explain its origin: is this molecular gas being stripped from the galaxy due to the ram-pressure, or is it formed {\it in situ} from the stripped neutral gas that has been able to cool down and is now prone to form new stars?
In order to answer this question, we further analyze the line-of-sight velocity of the molecular gas as revealed by the first moment of ALMA data.

\subsection{Molecular gas kinematics}\label{sec:kin}

From the first moment maps it can be seen that the two CO emission lines 
are broadly co-spatial (but the south-west tail show more concentrated CO(2-1) knots) and show similar velocities in the regions close to the galaxy disk.
It has to be noted, though, that first moments are intensity weighted, and therefore trace the motion of the summed components, in case of double peaked emission.
APEX spectra within the two broad A and B regions have shown that the molecular gas is rotating following the rotation of the ionized gas in the galaxy disk both in the central region and in the tail (hence the double peak emission), and ALMA data extracted from the same region confirm this results.

In order to better compare the cold molecular gas with the ionized gas motions, we constructed a velocity channel map by cutting the MUSE datacube around the \Ha emission \footnote{Unfortunately, the \Ha line is somewhat contaminated by the contiguous NII lines.} in the same velocity range covered by ALMA observations, and rebinning the ALMA data to match the MUSE velocity resolution (a dedicated cube with the same velocity resolution and pixel size of the MUSE data has been produced to this purpose).

\begin{figure*}
    \centering
    \includegraphics[width=0.9\textwidth]{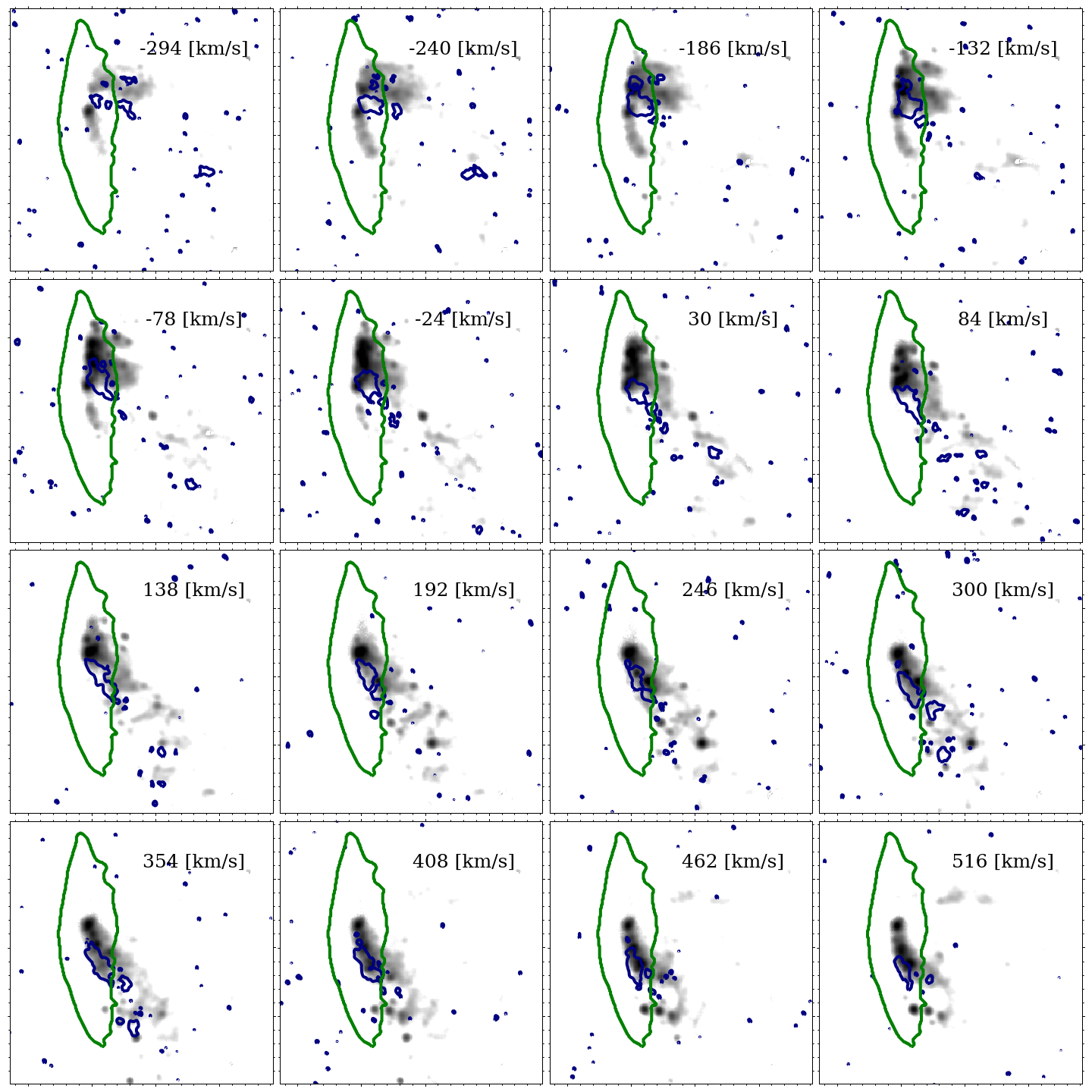}
    \caption{Velocity channel map showing the \Ha emission at different velocities, with superimposed in blue the contour of the CO(2-1) emission from ALMA data. The field of view of each plot corresponds to the one shown in the other figures of the paper.}
\label{fig:vchan}
\end{figure*}

Overall, we can see from Fig.~\ref{fig:vchan} that the molecular gas broadly follows the ionized gas emission, i.e. consistently with the \Ha velocity field, the molecular gas in the tail retains a memory of the disk rotation. However, as observed in \Ha, the CO also exhibits a decrease in the line-of-sight velocity as the distance from the disk increases westward. 

A comprehensive characterization of the galaxy kinematics is beyond the scope of this paper, and will be analyzed in a forthcoming study.
Here, however, we aim at understanding the broad characteristics of the cold gas behaviour as revealed by ALMA.

\begin{figure}
    \centering
    \includegraphics[width=0.45\textwidth]{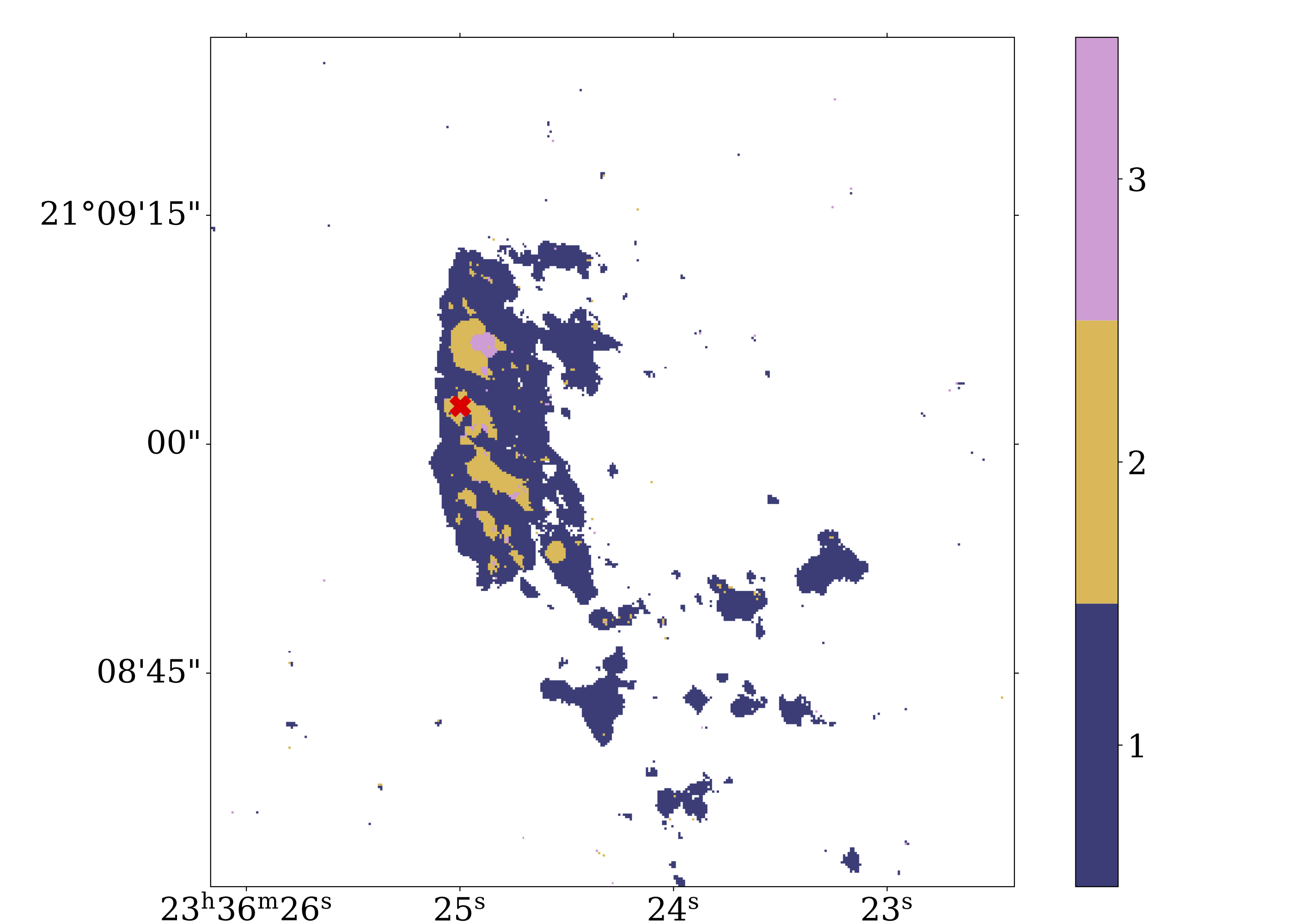}
    \caption{Number of gaussians needed to fit the CO(2-1) line emission from ALMA+ACA data. The red cross marks the AGN position.
    }
    \label{fig:2gfit}
\end{figure}

Given that the \Ha emission of JW100 shows a double component \citep{poggianti2017}, we tried to understand whether the molecular gas shows the same behaviour, by using a multi-gaussian fitting of the CO line emission on each pixel above the 3$\sigma$ level.
We have used the {\it gausspy+} code, that automatically evaluates the number of gaussians needed to fit a spectrum through subsequent refinements that take into account the presence of blended or broad components \citep{gausspy+}. 
We found that in some regions, mostly within the galaxy disk, the CO(2-1) line needs to be fitted using two gaussians, even with the ALMA enhanced spatial resolution (yellow regions in Fig.~\ref{fig:2gfit}). Fitting the CO(2-1) emission with only one gaussian
would produce artificially high velocity dispersion values (as can be seen in the lowest panels of Fig.~\ref{fig:Moments}).
The few spaxels where a third component is needed are not coincident with the AGN position (marked with a red cross).

The two components within the galaxy disk and in its vicinity (up to 1-2 kpc from it) are well explained if we think that one is associated with the gas rotating in the disk, and the second (the less dense one) is being stripped due to the ram-pressure.
The CO(2-1) emission in the tail regions can be fitted instead with one single gaussian, and can be either born {\it in situ}, or stripped as well. In this last case, though, we expect it to be characterized by a low density, as this component should be more easily stripped. We will try in Sec.~\ref{sec:r21} to assess this point through the estimate of the molecular gas density through the $r_{21}$ ratio.

The rotation of the molecular gas in the disk while ram-pressure is in action can also explain the accumulation of molecular gas in the D1 region, as all disk gas at the upstream edge that is pushed back by the ram pressure passes through it. If this is the case, also multiple kinematics components are not surprising.
The cold gas motion
could be in fact slowed down by the overall motion of the galaxy within the Intra Cluster Medium (ICM), which would result in accumulation of approaching (blue-shifted) gas if the galaxy is falling toward us and, therefore, the ram pressure wind component along the line of sight points away from us. The receding (red-shifted) gas, on the other side would be more easily removed, as the southern tail demonstrates.

Molecular gas outflows have been detected nowadays in many nearby active galaxies but they extend on smaller spatial scales, and show clear high velocity wings, that we can not see in our data  \citep{Cicone+2014,Feruglio+2015}. 
The northern D1 region, though, is far from the galaxy center ($\sim 5$ kpc) and it also has a different orientation with respect to the outflow 
shown by the ionized gas \citep{gaspXIX}, and we therefore think it is not due to a massive outflow.

\begin{figure*}
    \centering
    \includegraphics[width=0.45\textwidth]{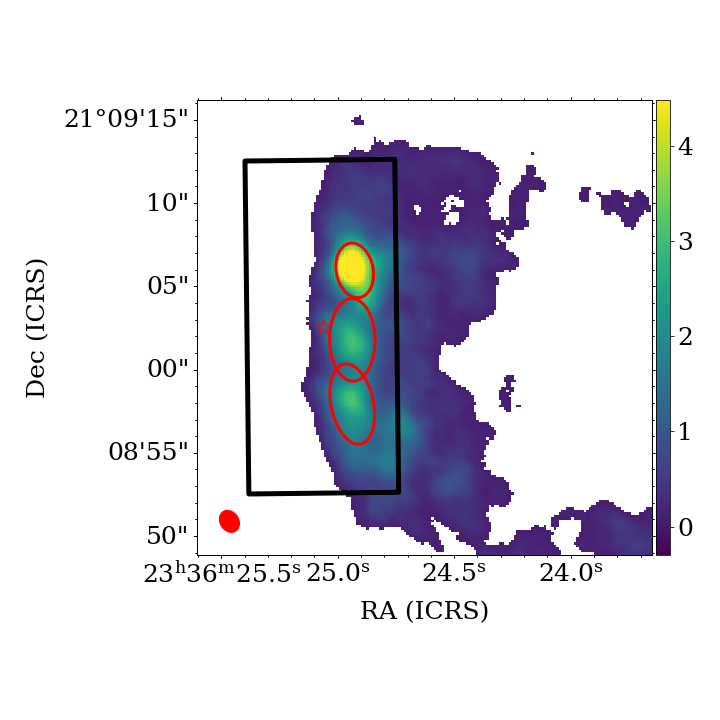}
    \includegraphics[width=0.45\textwidth]{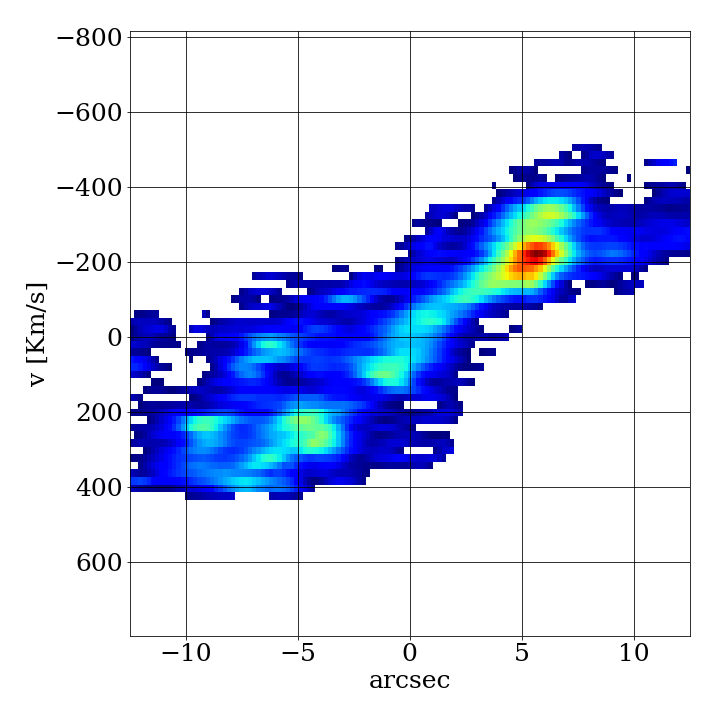}
    \caption{In the left panel the black box shows the $20 \times 9$ arcsec wide slit from which we extracted the Position-Velocity diagram shown in the right panel. Red ellipses show the position of the D1, D2 and D3 regions described in the text.
    }
    \label{fig:PV}
\end{figure*}

In order to understand the cold gas rotation we extracted from the masked CO(2-1) cube a generous slit 20 arcsec long and 9 arcsec wide, centered on the galaxy center, in the south-north direction with a Position Angle (PA) of 0.66 degrees (shown in Fig.~\ref{fig:PV}) and produced the position-velocity (PV) diagram shown in the right panel of the same Figure. 
We choose a wide slit width in order to include the bright CO emission west of the centre while still centering the slit on the galaxy itself. However, the PV diagram does not change significantly if we use an offset, narrower slit with the same PA and centred on the bright CO emission.

By comparing the left and right panel, it is easily seen that the CO distribution is concentrated in the three regions D1, D2 and D3 described above.

The PV diagram along the major axis of a nearly edge-on, massive galaxy should be characterized should be characterized by a broad S-shape (in velocity), due to the quickly rising rotation curve \citep[e. g.][]{Noordermeer+2007} and to the fact that at any given position along the extracted slit many components of the rotating disk should be intercepted. The rising part of the diagram should reach a maximum (at about 1 kpc) and then flatten. In our case, though, the gas shows a more linear behaviour, meaning that either some gas is missing from the central $\sim$ 3 kpc, as is the case if it is distributed within a ring, or that the rotation curve is still slowly rising at a radius of 3 kpc, which would be unusual for a galaxy of this mass.

It also suggests that there might be two different slopes, one tracing the rotation of the CO in regions D1 and D3, and a second steeper one in the central region D2, which also shows a discontinuity at both edges, possibly marking the extent of a ring. This possible secondary component could be due to the presence of a bar \citep{Funes+2002,Kuzio+2009}, and
the fact that the peak of the CO is displaced with respect to the center of this presumed bar is not totally unexpected \citep{Sorai+2000,Kuno+2007} and might trace a bar evolutionary sequence \citep{Sheth+2005,Jogee+2005}. 

As seen above, the CO kinematics shows two peaks in many regions within the galaxy disk.
This helps understanding the second moment maps, where the galaxy disk and region NT3 show a large velocity dispersion.
All the CO regions along the stripped tail have low velocity dispersions ($\leq 40$\kms), typical of star forming regions, as shown in the bottom panels of Fig.~\ref{fig:Moments}.

\section{H2 from different isotopes: deriving the R21}\label{sec:r21}

Given the complex environment in which JW100 resides, and in particular its position within an extremely disturbed cluster 
\citep{Gitti2013,Ignesti+2017,Ignesti+2018}, we can expect that the physical conditions among the detached knots of emitting CO will be different, as is different the surrounding ICM. Moreover,
as previously noted by \citet{gaspXIII}, most of the \Ha emitting knots detected with MUSE are characterized by optical line ratios typical of star forming regions, but some of them appear powered by shocks if using other line ratios.

Using ALMA data, and in particular the $r_{21}$ brightness temperature line ratio ($r_{21} \equiv T_{21}/T_{10}$, where the temperature comes from the CO emitting lines) 
we can now characterize the physical conditions of the molecular gas. 
This ratio is sensitive to both the density and the temperature structure of the gas, as well as the optical depth of the two lines.
Studies on the resolved Giant Molecular Clouds in Orion \citep{Nishimura+2015} have revealed that the inner portions of a GMC has usually a $r_{21}=1$ and then declines outward reaching $r_{21}\sim 0.5$.

\begin{figure}
    \centering
    \includegraphics[width=0.45\textwidth]{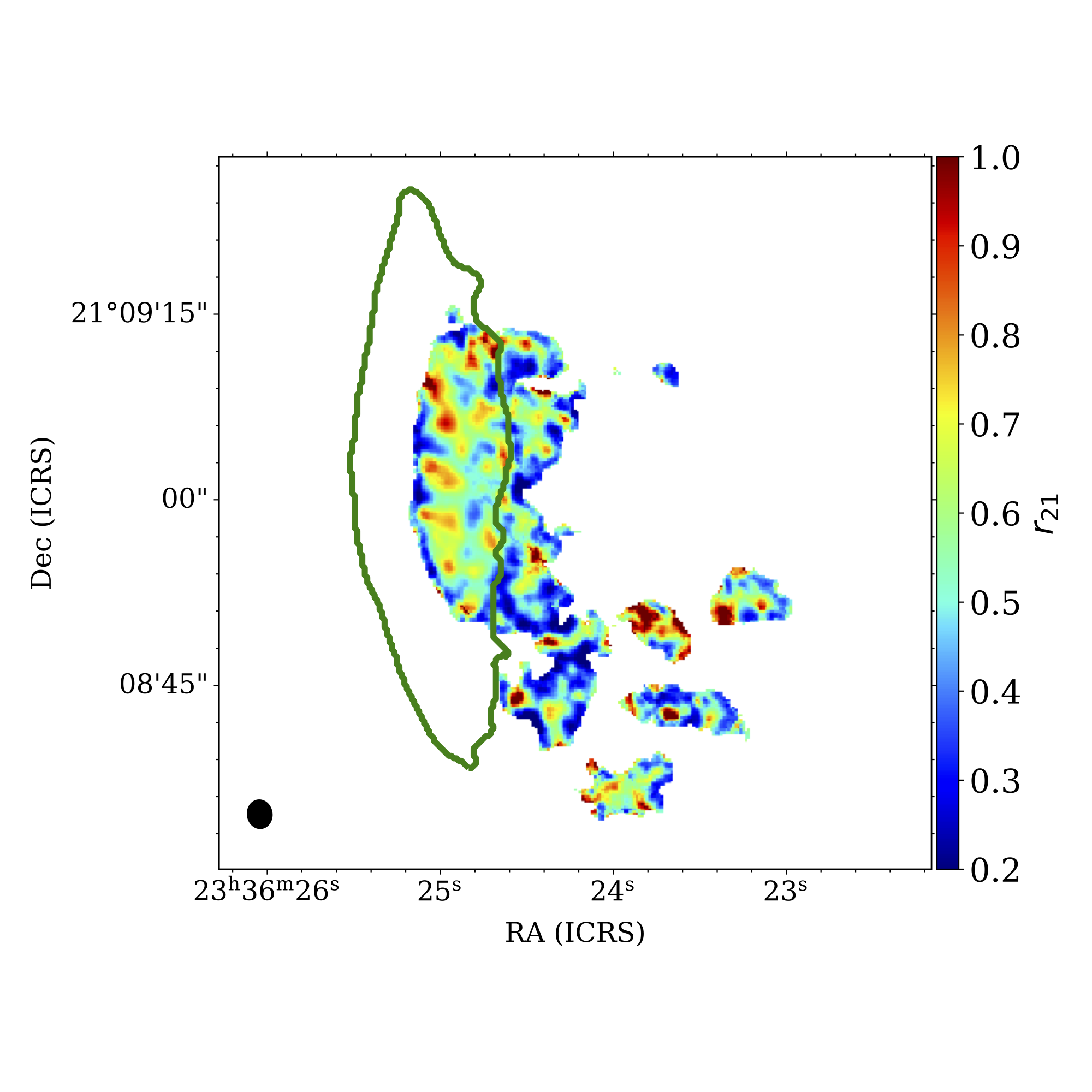}
    \includegraphics[width=0.45\textwidth]{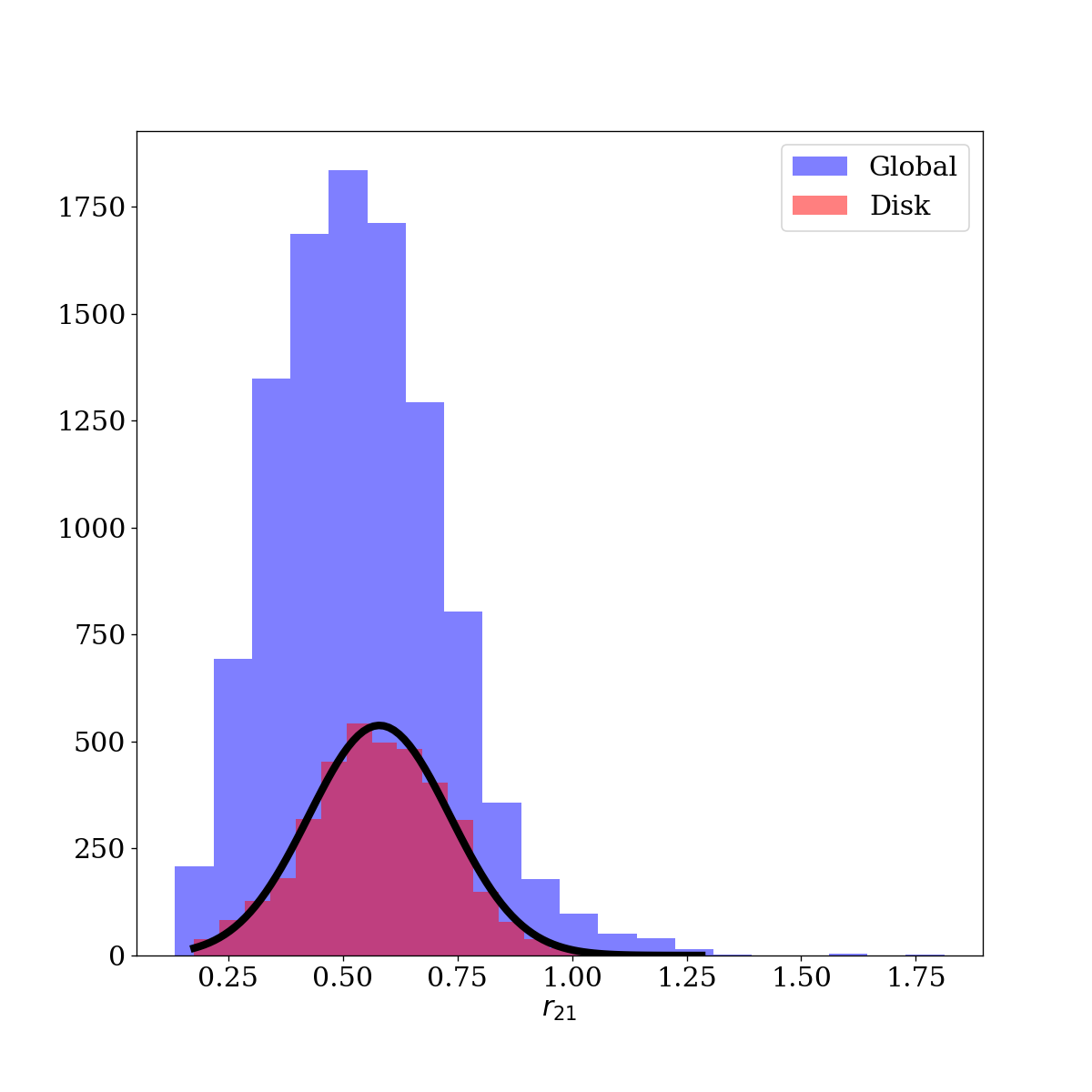}
    \caption{(top)$r_{21}$ ratio map with superimposed the galaxy stellar contour (in green).
    (bottom) $r_{21}$ distribution for the entire dataset (in blue) and for the disk only region (in red). The fit to the disk region is overplotted in black.
    }
    \label{fig:r21_map}
\end{figure}

We therefore constructed the $r_{21}$ map 
using the ALMA data as follows: 
we first masked the CO(2-1) and the CO(1-0) cubes at 20 \kms resolution, as described in Sec.~\ref{sec:moments}; then we smoothed the CO(2-1) datacube with a Gaussian kernel to get an image with the same beam of the CO(1-0) beam and regrid it on the CO(1-0) one; we finally extracted the ratio between the two  zero-moment maps (that gives the integrated brighntess temperature ratio, hereafter r$_{21}$ line ratio)
which is shown in Fig.~\ref{fig:r21_map}, top panel. Clearly, this value could be derived only when both measurements were available. The lower panel shows, instead, the distribution of the $r_{21}$ ratio, in blue for all the pixels where the measurement has been possible, in red for the disk region only. 

The $r_{21}$ distribution is very clumpy and goes from 0.1 to $\sim 1$ in the galaxy disk, while higher values ($\sim 1.5$) are found in the tail.

A gaussian fit of the distribution (shown in green in the lower panel of Fig.~\ref{fig:r21_map}) finds an average value of $\sim 0.58 \pm 0.15$ in the galaxy disk, which is lower than the usually adopted value of $\sim$0.8 \citep{Leroy2009,Saintonge2017} that we also used in the analysis of APEX data.
The value that we find is compatible with the one found in
nearby resolved studies of star forming disks \citep{Leroy2013}.
Extremely high values are found in galaxy nuclei, that can reach $r_{21}\sim1$, \citep{Leroy2009}, while
the lower tail of values at $\sim 0.3$ has been interpreted as due to optically thin, sub-thermally excited CO in warm (T$>$40K), diffuse ($n<10^3 cm^{-3}$) regions of single simulated clouds \citep{Penaloza+2017}.
Similar values have been also reported in the outskirts of the Orion region \citep{Nishimura+2015}.

In JW100, besides the low average value, the distribution appears very clumpy both in the disk and in the tail.
In particular, small regions of $r_{21}\geq 1$ are found in the stripped tail, but they are not spatially coincident with the peaks of the CO emission defining the regions described above, nor with the peaks of the \Ha emission.
As discussed in Poggianti et al., (accepted) this can be due to evolutionary effects related to the star formation process.

The estimated mean value found in every analyzed region is reported in col. 9 of Tab.~\ref{tab:regions} together with its error (standard deviation).

Our $r_{21}$ determinations confirm that within the galaxy disk the gas is optically thick and cold (with line ratios lower than 0.8 in the regions called D1, D2, and D3). Region NT1 is characterized by a particularly low value (0.45), that is thought to trace more diffuse and warm gas ($\sim 40$K), characterized by a very faint emission \citep{Penaloza+2017}. This CO complex lies in fact in the northern region around JW100, where the X-rays emitting gas is more concentrated (Poggianti et al., accepted), and might have had the chance to heat the molecular gas.
The $r_{21}$ increases in clumps located within the farther regions, implying that along the tail the gas becomes optically thick, being dense and warm. 
The stripping of such dense clumps out to a distance of $\sim 30$ kpc from the center of JW100 assuming the typical lifetime found in nearby massive galaxies, i.e. $\sim$5 Myr \citep{Chevance+2019}, would imply a stripping velocity of $\sim 5900$ \kms, which is rather unlikely.
Therefore, these dense clouds would be destroyed before reaching the position of the farthest clumps along the tail of JW100, supporting a scenario where molecular gas in the stripped tail is originated {\it in situ}. 

High values of the $r_{21}$ ratio are usually found in the central regions of galaxies \citep{Casoli+1991,BraineCombes1992} where new stars are born.
Instead, the D2 region in JW100, which is the closest to the galaxy center, has $r_{21}=0.62$, possibly because the energetic source here is the central AGN, and not SF.

\section{Quantifying the H2 mass} \label{sec:mass}

We now use the molecular gas emission (both the CO(2-1) and the CO(1-0)) to evaluate the total CO fluxes and then derive the molecular gas mass adopting the formulations by \citet{WatsonKoda2016}, i.e.

\begin{equation}\label{eqn:wk_co21}
\left(\frac{M_{\rm H_2}}{M_{\odot}}\right) = 
3.8 \times 10^3 \left( \frac{\alpha_{10}}{4.3}\right)
\left(\frac{r_{21}}{0.7}\right)^{-1}
\left(\int S_{21}dv \right)
\left(D_L \right)^2
\end{equation}

and

\begin{equation}
\left(\frac{M_{\rm H_2}}{M_{\odot}}\right) = 
1.1 \times 10^4 \left( \frac{\alpha_{10}}{4.3}\right)
\left(\int S_{10}dv \right)
\left(D_L \right)^2
\end{equation}

where 
$\rm \alpha_{10}$ is the CO-to-$\rm H_2$ conversion factor expressed in $M_{\odot}pc^{-2}$ \citep{Bolatto2013}, $r_{21}$ is the CO $J=2-1/1-0$ line ratio, $S_{21}$ is the CO integrated line flux in Jy and D$_L$ is the luminosity distance in Mpc.
Using this formulation or the one from \citet{Solomon2005} does not change more than 5\% our masses.
Throughout this paper we always use $\rm \alpha_{10}=4.3$, i.e. the standard Milky Way value corresponding to CO-to-$\rm H_2 = 2\times 10^{20}$ $cm^{-2}(K km s^{-1})^{-1}$ including the helium correction. As for the $r_{21}$, we will use the same value (0.79) that we used in our APEX data analysis for the sake of comparison.
We will also estimate the gas masses from the CO(2-1) adopting the observed value of $r_{21}$ and the masses from the CO(1-0).

A first question we can answer at this point is to what extent the overall \hdue mass is compatible with the one measured with the APEX single dish, as with ALMA we might be less sensitive to the largest scales of CO emission.
A mismatch between the interferometric flux and the one measured with the single dish is expected.
In fact, numerical simulations by \citet{Helfer+2002} find that ALMA would recover only $\sim 75\%$ of the total flux on large scales, this value being also dependent on the S/N and on the distance from the galaxy center, as confirmed by observations \citep{Helfer+2003}.

 In order to derive \hdue masses, we simply integrated the ALMA spectra within the two A and B regions defined by APEX pointings, and performed the same integral on the APEX spectra both within the same velocity range covered by ALMA detections.
The results are given in Tab.~\ref{tab:masses}.
Errors on ALMA fluxes amount to $\sim 10\%$, while those on APEX data are at least double \citep{Dumke2010}.

The percentage of flux loss when using the ALMA+ACA combination is similar in the two pointings ($\sim$10\%), and compatible with the measurements errors.
However, when considering the broader velocity range covered by the APEX spectrum in the pointing A, the comparison of the two fluxes reveals the presence (at about $\sim 1 \sigma$ level)  of a supplementary gas component at high velocity (between 500 and 800\kms) that is completely missed by ALMA and ACA observations, which could possibly suggest the presence of a diffuse components on large scales.

By integrating the spectra using the 12m array data only, we miss $\sim$43\% of the flux in the galaxy disk and $\sim$70\% in the tail region.

This means that both in the disk and in the tail the CO emission is more diffuse than the largest scale recovered by ALMA-12m, and that ACA is mandatory to recover the total flux.
As previously found by \cite{Pety+2013,Jachym+2019}, the molecular gas in the tail is therefore more diffuse than in the disk.

Estimating the total amount of molecular gas within this galaxy has deep consequences on the understanding of the origin of the gas itself: in fact, single dish surveys \citep{Saintonge2011}, as well as interferometric observations of nearby undisturbed galaxies from CALIFA \citep{Bolatto+2017}, give an estimate on the ratio of molecular over stellar mass which turns out to be $\sim 0.01$ (for stellar masses larger than $10^{11}$ \Msun), so that for JW100 the expected value of molecular gas is $\sim 3 \times 10^{9}M_{\odot}$.
The number of massive galaxies for which molecular gas masses are available in \citealt{Bolatto+2017} is very limited, but none of them reaches ratios larger than 0.02.
More massive galaxies are present in the COLDGASS sample \citep{Saintonge2011}, but again the molecular gas mass ratio never exceeds 6\% of the stellar mass.

We find that ALMA data account for a
total \hdue mass of 
 $\sim 2.5\times 10^{10}$\Msun.

If we consider as stellar disk the generous definition that is shown as green contours in all the maps, we end up with a total molecular gas mass in the galaxy disk of 
 $ 1.8\times 10^{10}$\Msun
, while in the tail it amounts to 
$0.7\times 10^{10}$\Msun. 

The disk value 
is compatible with the ones found by the HERACLES survey \citep{Leroy2009} of nearby undisturbed galaxies (which however
have lower stellar masses), but is
about one order of magnitude larger than the typical molecular gas mass found in galaxies of similar stellar mass in Virgo \citep{Corbelli+2012}.

Unfortunately, we could not observe the HI emission of this galaxy with the J-VLA, as done for other GASP jellyfishes \citep{gaspXVII,deb19}, due to the presence of radio interference at its redshift.
We are currently obtaining MeerKAT observations for this galaxy. 

Assuming for this galaxy a normal HI content, i.e. \hdue/HI=0.3 \citep{Saintonge2011}, we end up with a total gas fraction of $0.32\times$ the stellar mass,  which is 10 times larger than the expected value for galaxies with similar mass at the same redshift \citep{Catinella+2018}.
In fact, our \hdue mass estimate already provides a molecular-only gas mass fraction of  0.08, i.e. two times larger than the expected value, without accounting for the HI gas.
In any case, whatever the contribution of the HI gas, JW100 has a very high content of molecular gas.
Similarly high molecular gas fraction have been found also in NGC3627, a barred interacting and active (LINER) galaxy belonging to the NUGA survey \citep{casasola+2011} as in other interacting galaxies \citep{Kaneko+2017}.
Active galaxies with similar masses in the xCOLD GASS sample show, instead, lower molecular gas fractions, but they are not ram-pressure stripped/disturbed.

For each region defined in Sec.~\ref{sec:moments} we estimated the CO flux from the integrated spectrum.

Molecular gas in the near tail regions (NT1, NT2 and NT3) amounts to $1.13 \times 10^9$ solar masses, while the far tail clumps (FT1 to FT5) account for $\sim 0.8 \times 10^9$ solar masses of \hdue.
These values, though, refer to the small elliptical regions shown in Fig.~\ref{fig:map}
i.e. they do not take into account the more diffuse emission.
If we integrate our zero moment map, instead, we find that outside the disk $\sim 7  \times 10^{9}$ \Msun of molecular gas are present.

\begin{table}
\centering
\caption{CO(2-1) fluxes derived from the integrated spectra of ALMA data: col. 2 refers to the 12m observations only and col.3 to the combined interferometric and ACA data. Col. 4 refers to the integrated fluxes of APEX spectra from \cite{gaspX}
in the same velocity range of ALMA data.
}\label{tab:masses}
\begin{tabular}{llllll}
Region    & CO Flux  & CO Flux  & CO Flux  \\
          & Jy km/s  & Jy km/s  & Jy km/s  \\
\hline
&\multicolumn{1}{c}{12m}&\multicolumn{1}{c}{12m+ACA}&\multicolumn{1}{c}{APEX}\\
\hline
JW100 [A]   & 63   & 99  & 110 \\
JW100 [B]   & 11   & 32  & 36  \\
\hline
\end{tabular}
\end{table}

\begin{figure}
    \centering
    \includegraphics[width=0.45\textwidth]{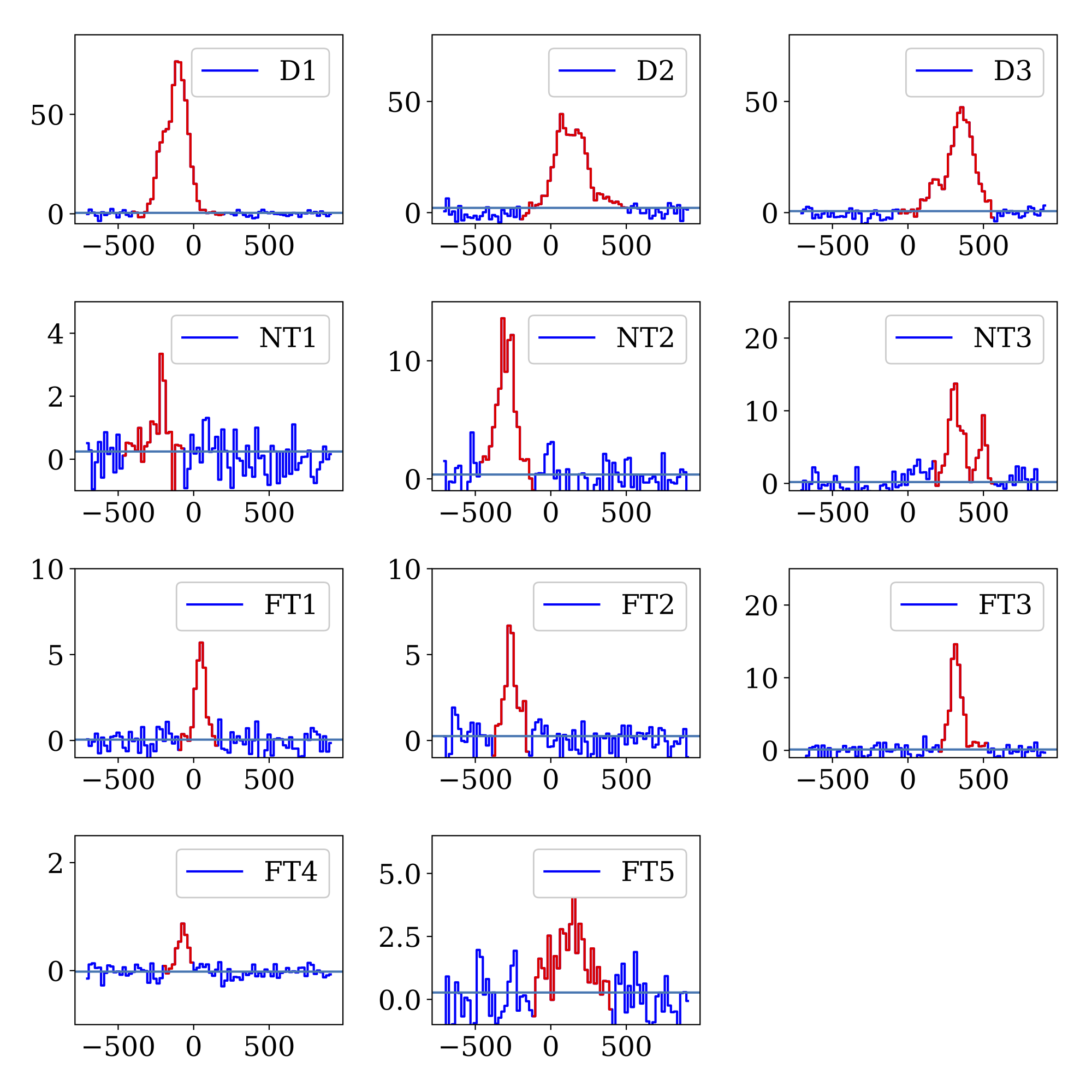}
    \caption{CO(2-1) integrated spectra in the selected regions D1 to FT5 defined in Sec. \ref{sec:moments}. The red line traces the flux integrated to obtain the molecular gas mass. Blue horizontal line is the median flux density within the region.}
    \label{fig:reg_spectra}
\end{figure}

We give in Tab.~\ref{tab:regions} the flux (and corresponding mass) of each region
while in Fig.~\ref{fig:reg_spectra} we show the CO (2-1) flux density within the selected regions.

For each region we estimated the molecular gas mass both from the CO(2-1) and from the CO(1-0) fluxes.
The two masses are in good agreement when using the appropriate value of $r_{21}$ (see Sec. \ref{sec:r21}) that we have obtained as the mean value within the analyzed region. The use of the fixed value of $r_{21}$, would provide an underestimation of the amount of molecular gas.

The strongest CO peak (region D1) contains 2.7-3.0 $\times 10^9$ \Msun of \hdue, while $\sim 80\%$ of this quantity is located in the region closest to the galaxy center (region D2). The amount of molecular gas slightly decreases going south (region D3).
It is interesting to note that the D2 region is not coincident with the AGN position (located at 23:36:25.0,+21:09:02.5 , from \citealt{gaspXIX}), meaning that the peak of the CO flux is displaced with respect to the active nucleus.
The three concentrations (regions NT1, NT2, and NT3) seen western from the galaxy disk have different contents of molecular gas: while the northern one (region NT1) has $4-7\times 10^8$ \Msun of \hdue, moving south the molecular gas mass increases, but these regions are intrinsically larger with respect to the NT1 region, where our 2D-fit found a very thin ellipse.
Within the largest CO emitting regions in the tail we identified with our procedure five emitting blobs (FT1 to FT5), with masses (within the ellipses) above $\sim 1\times 10^7$ and reaching $\sim 3 \times 10^8$ \Msun.
With our spatial and mass resolution, we can not say whether our complexes can be made of
Milky Way-like Giant Molecular Clouds \citep{Solomon+1987}.

\begin{table*}
\centering
\caption{CO(2-1) and CO(1-0) fluxes of selected regions (cols. 4 and 7), with the corresponding \hdue masses calculated using $r_{21}=0.79$ (col.5) and the measured one (col. 6). The $r_{21}$ measured from ALMA data is given in col. 9.}\label{tab:regions}
\begin{tabular}{lllllllll}
Region      & RA            &   DEC         & CO(2-1)  & $M(\rm H_2)$     & $M(\rm H_2)$     & CO(1-0)       & $M(\rm H_2)$   & r$_{21}$ \\
            &               &               & Jy km/s  & $10^9M_{\odot}$  & $10^9M_{\odot}$  & Jy km/s       & $10^9M_{\odot}$&          \\
            &               &               &          & $r_{21}=0.79$    & $r_{21}=meas$    &               &                &          \\     
\hline
JW100 D1       & 23:36:24.927  & +21:09:05.958 & 13.34    &     2.71        &   3.02        & 4.46 & 2.93  & 0.71 $\pm$ 0.11 \\
JW100 D2       & 23:36:24.937  & +21:09:01.784 & 10.57    &     2.15        &   2.74        & 3.94 & 2.62  & 0.62 $\pm$ 0.11 \\
JW100 D3       & 23:36:24.937  & +21:08:57.929 & 9.59     &     1.95        &   2.41        & 3.71 & 2.46  & 0.64 $\pm$ 0.08 \\
JW100 NT1      & 23:36:24.556  & +21:09:12.173 & 0.21     &     0.04        &   0.07        & 0.12 & 0.08  & 0.45 $\pm$ 0.17 \\
JW100 NT2      & 23:36:24.457  & +21:09:06.368 & 1.79     &     0.36        &   0.50        & 0.76 & 0.51  & 0.57 $\pm$ 0.09 \\
JW100 NT3      & 23:36:24.510  & +21:08:53.082 & 1.93     &     0.39        &   0.56        & 0.84 & 0.56  & 0.55 $\pm$ 0.12 \\
JW100 FT1      & 23:36:23.710  & +21:08:49.577 & 0.44     &     0.09        &   0.10        & 0.15 & 0.10  & 0.69 $\pm$ 0.13 \\
JW100 FT2      & 23:36:23.252  & +21:08:51.774 & 0.53     &     0.11        &   0.17        & 0.27 & 0.18  & 0.51 $\pm$ 0.11 \\
JW100 FT3      & 23:36:24.341  & +21:08:42.878 & 1.27     &     0.26        &   0.33        & 0.49 & 0.32  & 0.62 $\pm$ 0.10 \\
JW100 FT4      & 23:36:23.429  & +21:08:42.543 & 0.06     &     0.01        &   0.01        & 0.02 & 0.01  & 0.69 $\pm$ 0.06 \\
JW100 FT5      & 23:36:23.956  & +21:08:36.501 & 0.61     &     0.12        &   0.16        & 0.23 & 0.16  & 0.62 $\pm$ 0.09  \\

\end{tabular}

\end{table*}

\section{Resolved Star Formation Efficiency}\label{sec:sfe}

Having estimated the overall characteristics of the molecular gas in JW100, we now move on to the comparison of the ALMA and MUSE results, i.e. we correlate here the star formation measured through the \Ha emission from MUSE data and the mass of molecular gas.

\begin{figure*}
    \centering
    \includegraphics[width=0.95\textwidth]{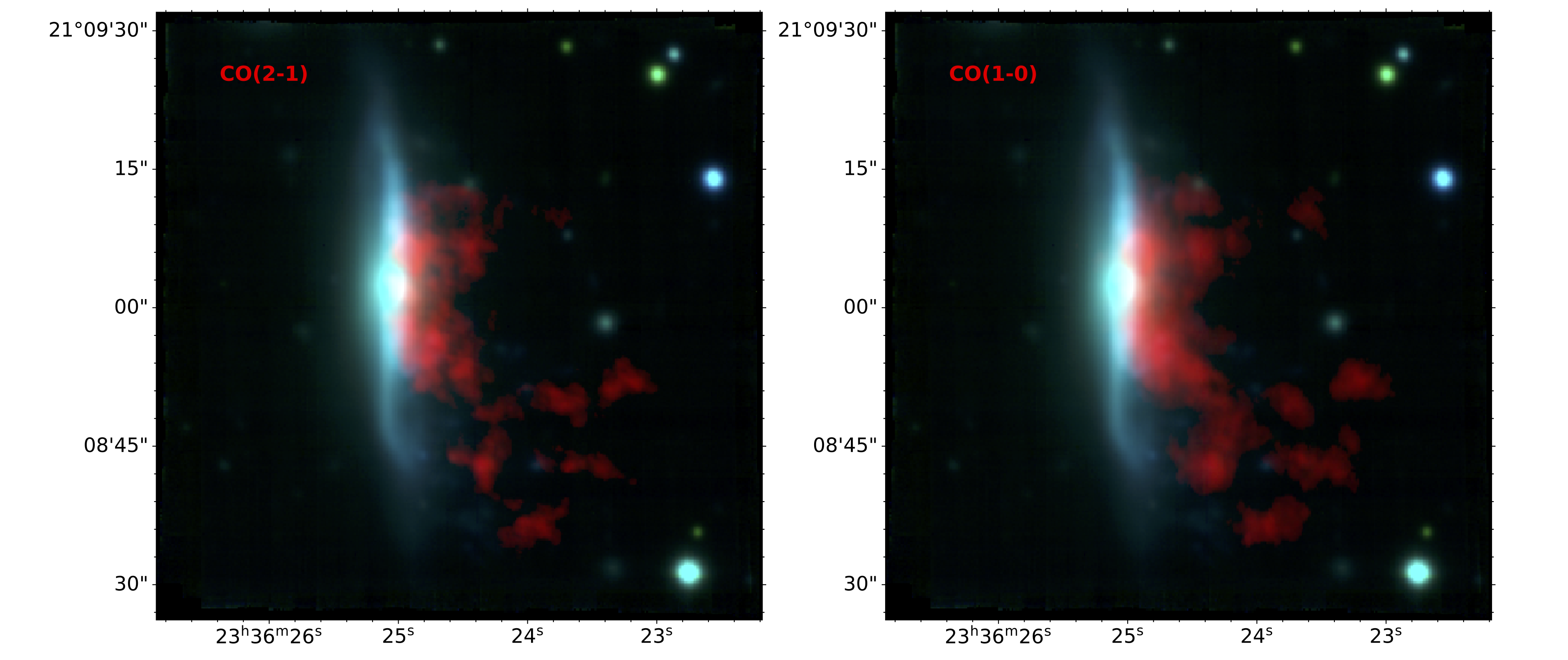}
    \caption{RGB images of JW100 obtained using the V (in blue) and I (in green) band images extracted from the MUSE datacube, and the CO emission as the red channel: left panel is made with the CO(2-1), right panel with the CO(1-0).}
    \label{fig:RGB}
\end{figure*}

Fig.~\ref{fig:RGB} shows the RGB images obtained by using the V and I band images extracted from the MUSE datacube and the CO zero moments from ALMA. 
The blue/green emission traces therefore the contribution of stars from MUSE, while the red regions show where the molecular gas is located.
In particular, it can be seen how the molecular gas distribution in the disk is restricted to a region of $\sim 20$ kpc, while stars are distributed over larger scales. Moreover, the CO is totally displaced towards the west side of the galaxy, as described in previous sections. Finally, it seems to avoid the very central region where there is a hint for the cavity/ring described in sec. \ref{sec:moments}.

In order to derive the Star Formation Efficiencies (SFE) over the 1 kpc scale covered by both the MUSE data and by the ALMA beam, we first convolve them to the same resolution and regrid them to the same WCS grid.

\begin{figure*}
    \centering
    \includegraphics[width=\textwidth]{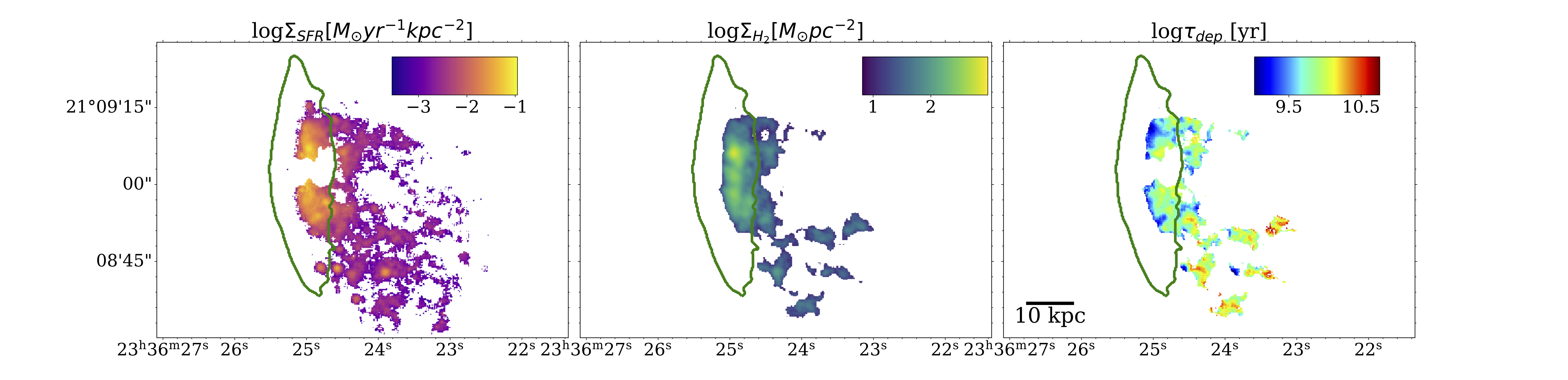}
    \includegraphics[width=0.95\textwidth]{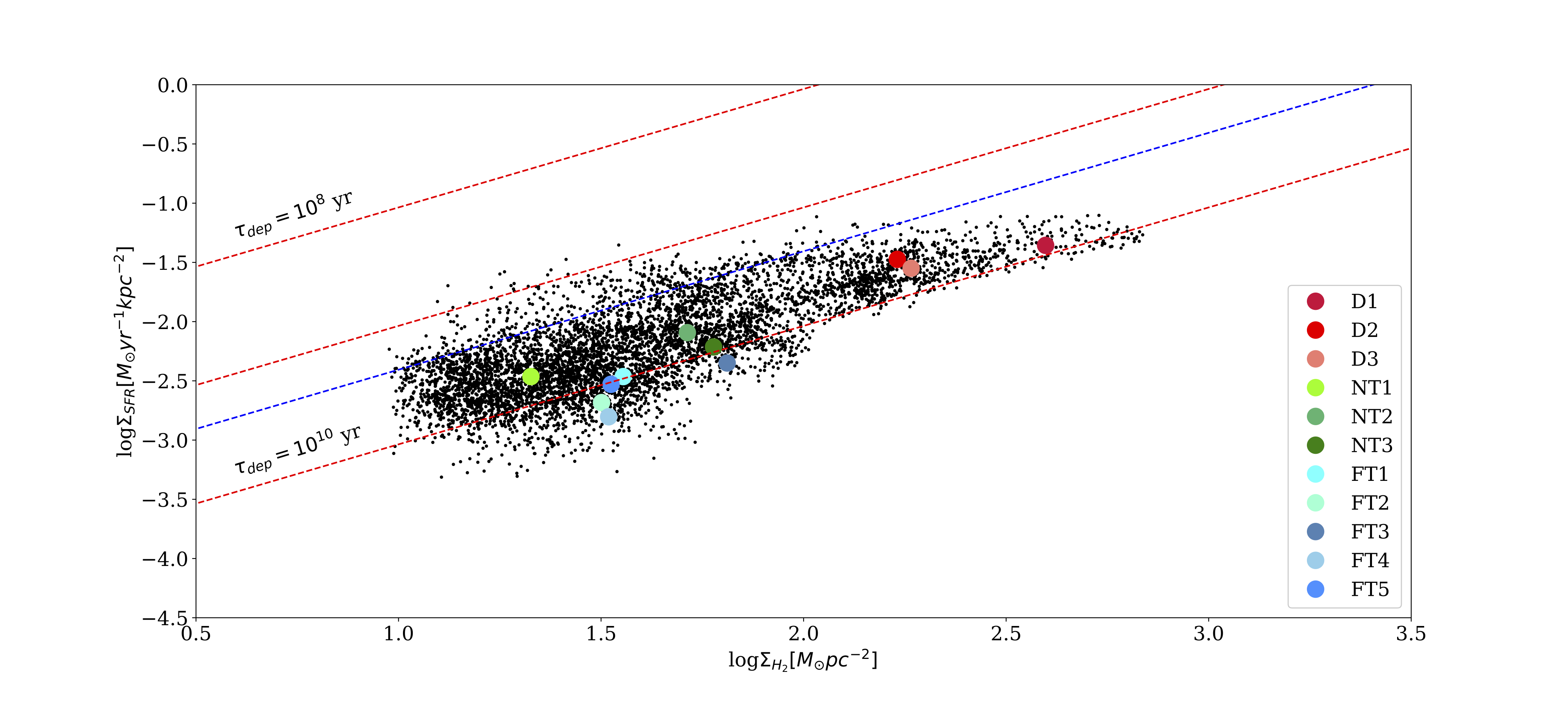}
    \caption{In the upper row: (left) Map of the SFR density of spaxels classified as star forming in MUSE (according to the NII lines); (middle) Map of the molecular gas mass density as derived from ALMA CO(2-1) data; (right) Map of the corresponding depletion time. In the lower row: Star Formation Rate densities against molecular gas mass densities pixel by pixel (in black) and averaged values for the analyzed region (colored symbols). The red dashed lines are fixed depletion times ($10^8$, $10^9$ and $10^{10}$ yr from top to bottom, respectively), while the blue dashed line is the average relation from \cite{Bigiel2011}.
    }
    \label{fig:tau_map}
\end{figure*}

As for MUSE results, we analyzed the spectra by measuring the emission line fluxes on the stellar continuum subtracted datacubes and corrected them for the dust contribution using the Balmer decrement \citep{gaspI}.
We then converted the \Ha flux in SFR adopting the \citealt{Chabrier2003} IMF using the relation given in \cite{gaspI}.
We then convolved the MUSE data (that have a 1\arcsec\, PSF) to the ALMA beam ($1.4\times 1.1$\arcsec\,, with PA=33 deg). This has allowed us to estimate the $\Sigma_{SFR}$ shown in the upper left panel of Fig.~\ref{fig:tau_map}, for the MUSE spaxels that have been classified as star forming according to the BPT diagram \citep{BPT} involving the NII line.

We then converted ALMA fluxes in mass densities, using eqn. \ref{eqn:wk_co21}, and then regrid them onto the MUSE frame conserving the flux.
This produces the $H_2$ mass density image shown in the upper middle panel of Fig.~\ref{fig:tau_map}. 

Given that the two frames are now convolved to the same beam/PSF and have the same pixel size, it is straightforward to construct the depletion time (i.e. the time needed to consume the available molecular gas at the given measured SFR) map, shown in the upper rightmost panel of Fig.~\ref{fig:tau_map}, where we have also superimposed the stellar contour, as in previous maps. 

Depletion times in JW100 range from $\sim 10^{9}$ yr in the eastern edge of the disk to $\sim 10^{10}$ yr in the tail, showing a clear gradient moving towards west.
The average depletion time in the disk regions (D1, D2 and D3) is 6.6 Gyr, and becomes longer in the narrow tail regions (NT1, NT2 and NT3), where it reaches 7.3 Gyr. The FT regions in the far tail have depletion times longer than the Hubble time (13.9 Gyr on average).

Interestingly enough, the depletion times are generally larger than the typical value of $\sim 2$ Gyr \citep{Bigiel2011} even in the central region, confirming that the ram pressure is influencing also the molecular gas located in the galaxy disk. The molecular gas velocity dispersion shown in Fig.~\ref{fig:Moments} is indeed high within the galaxy disk, with values that go from $\sim 60$ to more than 100 \kms, that are much higher than the typical values found in nearby galaxies \citep{Wisnioski+2012}.
Varying depletion times in disturbed galaxies are found in other nearby galaxies \citep{Tomicic+2018} on 0.5 kpc scales, and long depletion times (larger than 10 Gyr) are also common in the external part of disk galaxies \citep{Bigiel2010}. 

The lower panel of Fig.~\ref{fig:tau_map} shows the SFR density against the \hdue mass density for each pixel (black dots), and the average value found within each of the analyzed region as colored symbols. Red dashed lines are fixed depletion times, while the blue dashed is the one derived from the 30 nearby disk galaxies of the HERACLES survey by \cite{Bigiel2011} at 1 kpc scale resolution.
The red dots, corresponding to the regions D1, D2 and D3 located within the disk, lie below the local galaxies relation (in blue), i. e. they show low Star Formation Efficiency (SFE). This confirms that this galaxy is forming new stars in the disk at a very low rate (see also \citealt{vulcani+2018_sf}), given that the measured values of \sighdue are those typically found in galaxy bulges \citep{Fisher+2013}.
Lower values of $\alpha_{CO}$ (5-10 times lower than the Milky Way value adopted here) have been found in the central part of some HERACLES galaxies \citep{Sandstrom+2013}, but only in the central kpc, while the region D1, D2, and D3 spans a larger extent of the galaxy disk.
In particular, the central kpc is dominated by AGN-like line ratios \citep{poggianti2017,gaspXIX}, according to different indicators \citep{gaspXIII}, and we therefore excluded it from the calculation.
Our data confirm that local conditions play an important role in determining the SF process, as already shown in the nearby galaxy M51 \citep{bigiel+2016}, and in other 29 nearby galaxies where dense gas tracer were available \citep{Usero+2015}.
Ram pressure then works in JW100 by enhancing the gas density in the disk, and at the same time suppressing the global SFR, resulting in long depletion times both in the tail and in the galaxy disk (as already suggested in \citealt{gaspX}). 
Whether this is accompanied also by an enhancement of the molecular gas fraction, as in \cite{Nehlig+2016}, can not be stated yet. Ongoing HI observations with MeerKAT will shed light on this issue.
It is worth noticing, though, that in the GASP jellyfish galaxy JO206 the HI depletion time turned out to be shorter than expected \citep{gaspXVII}, showing the opposite trend with respect to the molecular gas in JW100.

In the region in which we detect the ring, we measure both high gas densities and low SFE, possibly due to the fact that molecular gas within that region is gravitationally unbound, and therefore less prone to become dense and form new stars \citep{Momose+2010,Sorai+2012,maeda+2018}.
It has been suggested, in fact, that high mass galaxies, or galaxies with high molecular gas fraction may have longer depletion times \citep{Leroy2013}.

\section{Conclusions}\label{sec:conclusions}

In this paper we presented the first results of an ongoing ALMA campaign devoted to studying the complex baryon cycle that leads to the quenching of the star formation in galaxies subject to ram-pressure stripping in the dense cluster environment.
Our targets, among which the JW100 galaxy here analyzed, are galaxies belonging to clusters at redshift $\sim 0.05$, and are part of the GASP project \citep{gaspI}.

JW100 is the first not nearby jellyfish galaxy for which we can confirm the presence of molecular gas out to large distances from the galaxy center thanks to the ALMA data over scales comparables to those of optical data (i.e. $\sim 1$ kpc).
Only recently ALMA has been used to confirm single dish results by \cite{Jachym+2019}, for the much more nearby jellyfish galaxy ESO137-001 in the Norma cluster.

Our ALMA data reveal the presence of  $\sim 2.5 \times 10^{10}$ \Msun of molecular gas (lower limit), i.e. 
 $\sim$8\% of the galaxy stellar mass, a value that is at least eight times the one found in the local xCOLD GASS sample \citep{Saintonge2017}  and in the Virgo cluster galaxies \citep{Corbelli+2012} for galaxies of similar mass.
The measured scatter in both samples can not explain our measurements.
JW100 is therefore extremely rich in molecular gas, suggesting that part of it is newly formed as a consequence of the gas stripping. In fact, $\sim 30\%$ of the molecular gas that we measure is located in the stripped tail and could be newly formed there from the stripped neutral gas or from the diffuse molecular gas.

In JW100 the CO emission resolved by ALMA-12m, combined with the ACA, is able to completely recover the APEX fluxes within the measurement errors when considering the same velocity range.

To understand on what scales the molecular gas is diffuse we remind here that
the largest recoverable scales of ALMA 12m array and ALMA+ACA are 7.5\arcsec and 18\arcsec, respectively (see Sec. \ref{sec:data}), that convert in $\sim 8$ and $\sim 19$ kpc at the cluster distance.
Our data, therefore, suggest that 57\% of the molecular gas in the disk is concentrated on scales up to 8 kpc, while only 30\% of it is in the same conditions in the tail, demonstrating that the cold gas in the tail is more diffuse than in the disk \citep{Pety+2013,Jachym+2019}.

The molecular gas within the galaxy disk is totally displaced with respect to the stellar component, demonstrating that the molecular gas reservoir of JW100 is disturbed by the ram-pressure exerted by the hot ICM on the galaxy while it falls toward the cluster center.
A clear double component in the CO emission is detected in the D1 northern disk region, where molecular gas is accumulated due to the geometry of the system with respect to the infalling galaxy.
Molecular gas is also found in the galaxy tail, broadly co-spatial with the ionized gas distribution, and with the same kinematics.

The total mass of cold gas in the tail is $\sim 7\times 10^{9}$ \Msun, and appears concentrated in large regions with sizes of few kpc.
The most massive regions outside the stellar disk are close to the disk itself ($\sim 1-2$ kpc), but the farthest CO emitting regions are located at more than 35 kpc from the galaxy center, a distance that again suggests the formation of molecular gas {\it in situ} from the stripped HI gas or diffuse molecular gas. 
This has been also suggested by the recent study by \cite{Jachym+2019} on the nearby jellyfish galaxy ESO137-001, on the basis of the CO(2-1) ALMA data.

Our dataset includes also ALMA band 3 data, i.e. the CO(1-0) line emission, that has allowed us to measure the line temperature ratio ($r_{21}$).
The $r_{21}$ ratio is within the expectations in the galaxy disk, where it shows values from 0.2 to 1, with an average value of 0.58 (lower than the usual adopted value of 0.7-0.8). The regions located well outside the galaxy disk tend, instead, to show higher values. 
The high values of the $r_{21}$ ratio in the farther FT regions, together with their low velocity dispersion and the presence of a single emission line component lead us to conclude that they are composed by dense star forming molecular gas. The comparison between the typical lifetime of a dense molecular gas cloud and the time needed to reach such distances along the stripped tail strongly support the scenario in which these clouds are formed {\it in situ}, either from stripped neutral gas, or from stripped diffuse molecular gas.
The nearby clumps, instead, might bear the trace of a secondary, less dense, molecular gas component that has been stripped from the disk (see the narrow tail NT3 region for example).

The total amount of molecular gas, coupled with its distribution, kinematics and physical conditions are all suggesting that the molecular gas content of this galaxy is increased with respect to undisturbed galaxies of similar mass, and that this is probably due to the newly formed gas that we see in the tail coupled with an enhancement of molecular gas also within the disk.

By coupling the ALMA data and the GASP results obtained from MUSE spectra, we finally derive the SFE over scales of 1 kpc, for the first time in a distant jellyfish galaxy. 
We find that there is a clear gradient in the depletion time, with average values that go from  $\sim 6$ Gyr within the disk to $\sim 14$ Gyr in the farthest tail regions (FT1 to FT5). Most of the molecular gas in the tail, therefore, will not be used to fuel the SF, but will ultimately join the ICM.
The high value that we find even within the galaxy disk can be explained by an increased turbulence caused either by the bar or by the ram-pressure itself, that would make the gas unable to adequately form stars. This mechanism has been also proposed to explain the star formation quenching at the center of our own Galaxy \citep{Haywood+2016}, and in various regions in M51 \citep{querejeta+2019}, where the star formation efficiency has been found related to turbulent motions and galactic dynamics.
Both dedicated simulations of gas rich galaxies \citep{khoperskov+2018} and isolated galaxies with central spheroids with stellar mass densities larger than $\sim 3\times10^8$ \Msun \citep{gensior19} have confirmed this result . 
The comparison between ALMA-12m and APEX data also suggests that a significant amount of CO is diffuse on scales larger than the one recoverable with ALMA-12m, implying that the molecular gas densities might be even higher (especially in the tail), making the SFE consistently lower. 

Further analysis on other three GASP galaxies for which ALMA data are already available (JO201, JO204, JO206) will help in clarifying the effect of the ram-pressure stripping on this issue.

\acknowledgments
{We thank the anonymous Referee for the very constructive report that helped us to improve the paper. We acknowledge funding from the INAF PRIN-SKA 2017 program 1.05.01.88.04 and from the agreement ASI-INAF n.2017-14-H.0, as well as from the INAF main-stream funding programme.
B.~V. and M.~G. also acknowledge the Italian PRIN-Miur 2017 (PI A. Cimatti).
This project has received funding from the European Research Council (ERC) under the European Union's Horizon 2020 research and innovation programme (grant agreement No. 833824, GASP project and grant agreement No. 679627, FORNAX project).
Y.J. acknowledges financial support from CONICYT PAI (Concurso Nacional de Inserci\'on en la Academia 2017) No. 79170132 and FONDECYT Iniciaci\'on 2018 No. 11180558
Nacional de Inserci\'on en la Academia 2017) No. 79170132.
This paper makes use of the following ALMA data: ADS/JAO.ALMA\#2017.1.00496.S. ALMA is a partnership of ESO (representing its member states), NSF (USA) and NINS (Japan), together with NRC (Canada) and NSC and ASIAA (Taiwan), in cooperation with the Republic of Chile. The Joint ALMA Observatory is operated by ESO, AUI/NRAO and NAOJ. 
This research made use of APLpy, an open-source plotting package for Python (Robitaille and Bressert, 2012; Robitaille, 2019).
}

\vspace{5mm}
\facilities{ALMA, VLT(MUSE)}

\software{Astropy \citep{astropy,astropy2},  
          CASA \citep{CASA}
          }

\bibliography{references}{}
\bibliographystyle{aasjournal}

\end{document}